\documentclass[useAMS,usenatbib]{mn2e}
\usepackage{graphicx}
\usepackage{tabularx}

\title[]{Stellar rotation in the Hyades and Praesepe: gyrochronology
  and braking timescale}
\author[P. Delorme et al]{
P. Delorme$^{1}$\thanks{E-mail:pd10@st-andrews.ac.uk},
A. Collier Cameron$^{1}$,
L. Hebb$^{2}$,
J. Rostron$^{1}$,
T. A. Lister$^{3}$,
\newauthor
A. J. Norton$^{4}$,
D. Pollacco$^{5}$,
and
R. G. West$^{6}$\\
$^{1}$SUPA, School of Physics and Astronomy, University of St Andrews, North Haugh, St Andrews, Fife KY16 9SS, UK.\\
$^{2}$ Department of Physics and Astronomy, Vanderbilt University,  6301 Stevenson Center, Nashville, TN 37235, USA.\\
$^{3}$Las Cumbres Observatory, 6740 Cortona Dr. Suite 102,
Santa Barbara, CA 93117, USA.\\
$^{4}$ Department of Physics and Astronomy, The Open University, Milton Keynes, MK7 6AA, UK.\\
$^{5}$Astrophysics Research Centre, School of Mathematics \&\ Physics, Queen's University, University Road, Belfast, BT7 1NN, UK.\\
$^{6}$Department of Physics and Astronomy, University of Leicester, Leicester, LE1 7RH, UK.\\
} 

\begin{document}

\date{}

\pagerange{\pageref{firstpage}--\pageref{lastpage}} \pubyear{2010}

\maketitle

\label{firstpage}

\begin{abstract}
  We present the results of photometric surveys for stellar rotation
  in the Hyades and in Praesepe, using data obtained as part of the
  SuperWASP exoplanetary transit-search programme. We determined
  accurate rotation periods for more than 120 sources whose cluster
  membership was confirmed by common proper motion and
  colour-magnitude fits to the clusters' isochrones. This allowed us to
  determine the effect of magnetic braking on a wide range of
  spectral types for expected ages of $\sim$600Myr for the Hyades and
  Praesepe.  Both clusters show a tight and nearly linear relation
  between $J-Ks$ colour and rotation period in the F,G and K spectral
  range. This confirms that loss of angular momentum was significant
  enough that stars with strongly different initial rotation rates have
  converged to the same rotation period for a given mass, by
  the age of Hyades and Praesepe. In the case of the Hyades our
  colour-period sequence extends well into the M dwarf regime and
  shows a steep increase in the scatter of the colour-period relation, with
  identification of numerous rapid rotators from  
  $\sim$0.5M$\odot$ down to the lowest masses probed by our survey
($\sim$0.25M$\odot$). 
This provides crucial constraints on the rotational braking timescales
and further clears the way to use gyrochronology as an accurate age
measurement tool for main-sequence stars. 
\end{abstract}


\section{Introduction}
The coeval populations of stars in galactic open clusters provide the best
laboratories we have for testing theories of stellar evolution. The understanding we
have gained from modelling their colour-magnitude diagrams allows us to determine
the ages of open-cluster populations with far greater confidence than is possible 
for any individual field star. Following the establishment of an age sequence for 
open clusters, studies of stellar rotation by \citet{Kraft.1967}, \citet{Skumanich.1972}
and others revealed that the rotation rates of stars in open clusters
decline with cluster age. The power-law spindown with age $\Omega\propto t^{-1/2}$
is consistent with expectations from idealised models of angular-momentum loss 
via the stellar equivalent of the hot, magnetically-channelled solar wind \citep{Weber.1967}.

Photometric studies of rapid rotators in young open clusters flourished in the late 1980s, but
precise measures of the spin periods among the slower rotators and in older clusters remained elusive.
\citet{Radick.1987} found that main-sequence  F, G and K  in the Hyades
show a far tighter period-colour relation than is found in the younger populations
of the $\alpha$ Persei and Pleiades clusters. Models of spin-up and spin-down
in the pre- and post  zero-age main sequence
\citep[e.g. ][]{Cameron.1994,Cameron.1995,Bouvier.1997} 
provided further confirmation that power-law spindown should cause the
spin rates of otherwise  
identical stars with different rotational histories to converge within a few 
hundred Myr of the ZAMS.  These models employed the Weber-Davis
  formulation modified to take into account the dependance of the
  magnetic flux in the wind on 
rotation rate for the most rapidly spinning stars.
As \citet{Barnes.2003} pointed out,
this suggests a single-valued period-age relation for stars of a given mass 
older than the threshold age at which convergence
occurs.  \citet{Barnes.2007} provided the first attempts at
calibrating stellar spindown as a clock, using asteroseismology and
open clusters as primary age calibrators, and \citet {Mamajek.2008}
proposed a modified age-rotation calibration.

The study of stellar rotation in open clusters has undergone something 
of a renaissance in the last few years, largely as a by-product of
wide-field  
searches for transiting exoplanets in open clusters and in the field. The 
wide fields of view, superb photometric precision and long
durations of these surveys are perfectly suited to determining the rotation
distributions of young and intermediate-age open clusters with unprecedented 
completeness.  Many such studies focusing on cluster of various ages
are cited in the recent review by
\citet{Irwin.2009}. Since then, further studies of Coma Berenices \citep[][]{Cameron.2009},
NGC 2301 \citep[][]{Sukhbold.2009}, M34 \citep[][]{James.2010sub},
M35 \citep[][]{Meibom.2009_M35} and the 
Pleiades \citep{Hartman.2010sub} have been published. 

As two of the closest \citep[respectively 46pc and 180pc,][]{Vanleeuwen.2009} rich, intermediate-age open clusters to the Earth,  the Hyades 
and Praesepe provide a vital calibration points for many types of stellar evolution study. 
Photometry, however, presents a severe challenge, owing to the wide angular extent
of both clusters. Fortuitously, they both fall within the range of declinations 
surveyed by the northern camera array of the Wide-Angle Search for Planets (SuperWASP).
Here we present studies of the distribution of rotation periods as a function of colour 
in both clusters, derived from SuperWASP photometry. For the first time we extend the 
Hyades period-colour relation from the F, G and K stars studied by Radick et al, down to 
the late-K and M-dwarf population in which the magnetic braking process has not
yet led to convergence.


After presenting in section 2 the SuperWASP data we used for this
study, we explain in section 3 how we obtained the rotational periods
of 70 stars in the Hyades cluster. Section 4 focuses on the determination of the
period-colour relation in the Hyades and in section 5 we extend this
work and apply it to the Praesepe cluster. Section 6 is dedicated to the
comparison between the Hyades, Praesepe and Coma, with the
determination of their relatives ages. We use this age information,
together with the rotational data to put strong constraints on the
magnetic braking timescales. The last section deals with the
calibration of gyrochronology on the handful of stars with robust age
measurement to derive the age of F, G, K and early-M field stars from
their rotation periods.

\section[]{The data}
\subsection{Observations}
We determined stellar rotation rates using data from the SuperWASP camera array, located at the Observatorio del
Roque de los Muchachos on La Palma, Canary Islands. With its 8
cameras, SuperWASP \citep[described in detail in][]{Pollacco.2006} has
a total field of view of 640 deg$^2$. This extremely wide field of
view provides the ability to make repeated observations of areas of the
sky as large as the full Hyades cluster \citep[10deg radius,
according to][]{Perryman.1998}, providing densely sampled
photometric data. We used fields within $\sim$15deg of the Hyades
centre (04$^{\rm h}$ 26$^{\rm m}$.9 +15$^\circ$ 52$'$) with a good 
enough sampling and a large enough time base (more than 56 days) 
to efficiently probe rotation periods from about a
day up to about 20 days  without significant field to field
  bias.  The corresponding fields, detailed in 
table \ref{fields}, were observed between 25 and 100 times per night on
average, with typically 5 to 10 minutes between each 30s
exposure. The most sparsely-observed field, (0420+1410) was observed on
  25 usable nights spanning a baseline of 56 nights in the 2008
  season. This field yielded 8 candidates, two of which 
yielded similar periods in another season. Given the dense sampling (more
than 80 points per usable night) we do not expect any bias against short
periods in this field. The shorter baseline may decrease our completeness
at the long-period end of the search window. Such a bias cannot be strong
because we find many variable sources with periods $>$ 18 days among the
non-cluster stars in this field.
Note that these fields refer to individual camera exposures
and cover 62deg$^2$. Since there are overlaps between them, some
of the sources have been observed in different fields during different
observing seasons.

\begin{table}
 \centering
 \caption{Hyades SuperWASP fields used in this survey.\label{fields}}
\begin{tabular}{|c|c|c|c|c|c|c|} \hline
Field centre & No. of & Usable & First  & Baseline\\ 
             & images & nights & night & (nights) \\ \hline
03 16' +24 10' & 3078 & 58 & 2006-09-08 &135 \\ \hline
03 17' +23 26' & 1919 & 61 & 2004-07-06 & 85 \\ \hline
03 42' +24 32' & 2527 & 56 & 2006-09-16 &131 \\ \hline
03 46' +13 14' & 4721 & 51 & 2008-10-03 &113 \\ \hline
03 48' +06 00' & 4779 & 51 & 2008-10-03 &113 \\ \hline
03 51' +17 42' & 2535 & 56 & 2006-09-16 &131 \\ \hline
04 12' +17 34' & 2494 & 55 & 2006-09-16 &131 \\ \hline
04 16' +24 10' & 2555 & 56 & 2006-09-16	&131 \\ \hline
04 17' +00 44' & 4812 & 51 & 2008-10-03	&113 \\ \hline
04 17' +23 26' & 1382 & 47 & 2004-07-28 & 64 \\ \hline
04 18' +07 59' & 4808 & 51 & 2008-10-03 &113 \\ \hline
04 20' +14 10' & 2181 & 25 & 2008-11-06 &56  \\ \hline
04 42' +24 31' & 2759 & 54 & 2006-09-29 &140 \\ \hline
04 46' +13 14' & 5428 & 52 & 2008-10-12 &112 \\ \hline
04 48' +06 00' & 5437 & 52 & 2008-10-12 &112 \\ \hline
04 51' +17 41' & 2733 & 54 & 2006-09-29 &140 \\ \hline
05 12' +17 34' & 2697 & 53 & 2006-09-29 &140 \\ \hline
05 18' +07 59' & 5472 & 52 & 2008-10-12 &112 \\ \hline
05 18' +36 25' & 2725 & 54 & 2006-09-29 &140 \\ \hline
05 20' +14 09' & 4462 & 44 & 2008-11-06 &87  \\ \hline
05 22' +30 00' & 2739 & 54 & 2006-09-29 &140 \\ \hline	
\end{tabular}
\end{table}

\subsection{Reduction and calibration}
We used data reduced by the standard WASP pipeline,
described in detail by \citet{Pollacco.2006,Cameron.2009}. Each
SuperWASP source is matched with NOMAD and Hipparcos data, providing
high signal to noise photometry in the standard $BVJHK$ bands, proper
motion as well as parallax when available. SuperWASP photometric data
itself is only used to look for photometric rotational modulation of
the signal. Its long term rms scatter is 0.004 mag for a V=9.5
star, degrading to 0.01mag for a V=12 star
\citep[][hereafter CC09]{Cameron.2009}. This 
is achieved after a careful removal of the patterns of correlated
errors, using {\sc sysrem} \citep[][]{Tamuz.2005} as described in
\citet{Cameron.2006}.  The dominant systematics are secondary
extinction and temperature-dependant variations of camera focus
through the night. None of these have a timescale similar to the
rotational modulation expected for 600Myr old F, G, K or M stars.
We are therefore confident that the deconvolution process does not
induce spurious detections or remove genuine ones in the range of
interest. A more significant bias is linked to the lunar cycle which
seems, in the case of Praesepe (see figure \ref{P_initial_sample}), to
dramatically increase the number of periods identified around 19.5 days,
at a frequency equivalent to 2/3 times the lunar synodic period. As
described below, this period range was removed from our sample.
 Saturation of
 SuperWASP images at $V\sim$8.5  causes a progressive lack of periods
 for bright sources. At Hyades distance this translates into a biased
 lack of  blue members with $J-K_s<0.4$.  At the red end, only the
 closest Hyades members are detected in $V$-band with enough signal to
derive period. This cause a bias toward detecting only red objects brighter
than the Hyades main sequence at the centre of the cluster and
multiple star systems; see Fig. \ref{H_CMD}.

\section[]{Candidate selection}
\subsection{Light curve analysis \label{light_curve}}
\begin{figure}
 \includegraphics[width=6.25cm,angle=270]{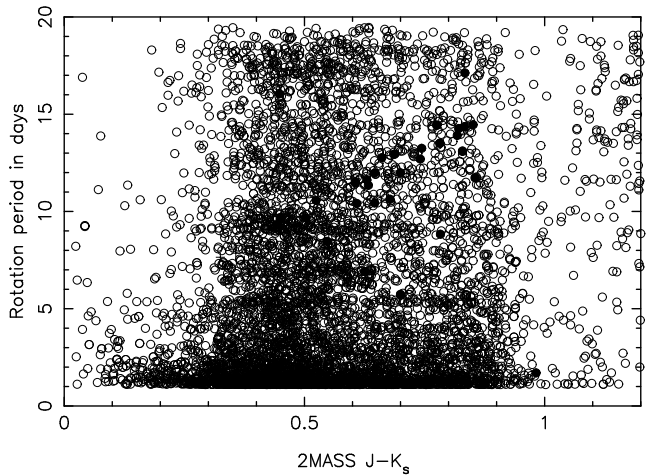}
 \caption{All sources with a good rotational signal (FAP$<0.05$)
   detected in the fields around the Hyades cluster. \label{initial_sample}}
\end{figure}

\begin{figure}
 \includegraphics[height=6.25cm,angle=0]{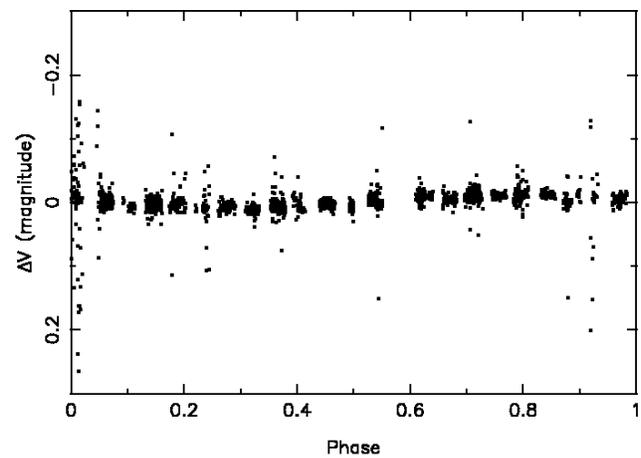}
 \caption{Folded light curve of one representative periodic variable
   source in the Hyades, 1SWASP J040339.03+192718.1. \label{folded}}
\end{figure}

 We used the generalised Lomb-Scargle
periodogram formulation of \citet{Zechmeister.2009} to look for
quasi-sinusoidal light curve modulation in all stars in the fields
detailed in table \ref{fields}. The details of the frequency analysis
and the optimised False Alarm Probability (hereafter FAP) calculation
we used are described by CC09. We looked for signal
modulation due to rotation
periods  between 1.1days and 20 days and selected only signals with a
FAP$<$0.05. We  did not look for rotation periods below 1.1 day because
for such short period most of the detections were caused by
observational artefacts around 1 day, 1 night and the associated higher
frequency harmonics.  Lower frequency harmonics of 1 day at periods of a few
days, though much less likely (see Fig \ref{initial_sample}), are possible
and we closely examined the light curves of all selected members
with periods close to n days, to remove such artefacts. Rotation periods above 20 days were not
investigated because our baselines were not always long enough to
efficiently probe this frequency range. Moreover, very few, if any, periods over
20 days are expected in $\sim$ 600Myr old clusters such as the
Hyades. The 21 Hyades SuperWASP fields yielded 40925 
periodically variable sources out of about 1~000~000 objects. As an
example, we show the folded light curve of one of these variable
sources in Figure \ref{folded}.  Since many of these fields overlap,
some variable sources have several
independently determined periods from multiple observations at
different seasons. 
 
\subsection{Cluster membership} 
Each source with a detected period was examinated to
determine whether or not it was a cluster member. 
A cluster membership probability was derived by comparing the proper
 motion and the apparent magnitudes of each candidate to those
 expected for Hyades members. 

\subsubsection{Proper motion} Given the  wide spatial extent of the Hyades on
the sky \citep[up to 20 degrees][, hereafter P98]{Perryman.1998} and its
proximity (46pc, P98) it is necessary to determine the motion of each
source with respect to the cluster convergent point instead of dealing
directly with RA/DEC proper motion. We therefore converted the proper
motion to a component going toward the convergent point ($\mu_{tan}$) and a
perpendicular component($\mu_r$), following the method described by
\citet{Reid.1992}. We could then robustly estimate how the proper motion of
each variable source differs from the Hyades 2-dimensional
space-velocity. We adopted a mean proper motion of the Hyades of $\mu_{tan}=-110\pm 0.017$ mas yr$^{-1}$ and $\mu_r=0.005\pm 0.006$ mas yr$^{-1}$ 
following \citet{Reid.1992}.\\

\begin{figure}
 \includegraphics[width=6.25cm,angle=270]{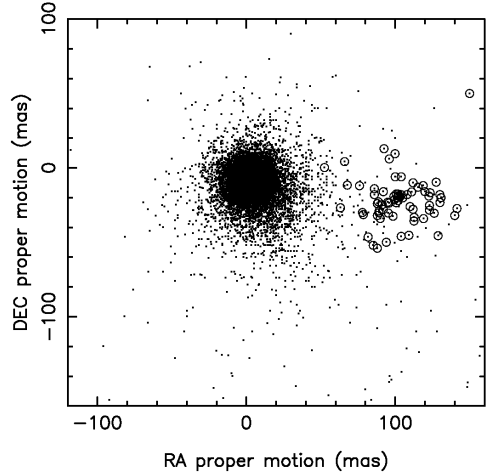}
 \caption{Proper motion in RA and DEC of all sources with identified rotational
   modulation signal in our survey. Objects identified as Hyades
   members owing to their proper motion and apparent magnitude are circled.\label{PM_Hyades}}
\end{figure}

The next step was to compute the average proper motion of fields
stars, to be able to estimate how the proper motion of
each variable source differs from the field 2-dimensional
space-velocity distribution. We
used a 3$\sigma$ clipped-average to eliminate
Hyades stars and other extreme proper motion outliers from the field
average and dispersion determination. 
Since Hyades stars make up for
less than 0.1\% of the sources and populate a significantly different
proper motion range (see figure \ref{PM_Hyades}) this effectively
removed them from the calculation.  The field proper motion
distribution was determined by computing the average and dispersion of
all remaining sources.   
The last step is to compare the relative proximity of each source to
the Hyades ($f_c$) and to the field ($f_f$) proper motion distributions. This was carried
out assuming Gaussian distribution of both population and weighted by the
relative number of stars in the field, $N_f$, and in the cluster
, $N_c$, to take into account that there are many more field stars
than cluster stars in our sample. $N_c$ was determined
by running the full membership routine recursively until $N_c$,
defined as the number of sources with a cluster membership probability over
50\%, converged. $N_f$ is the size of the sample minus $N_c$. The
related equations are the following:\\
\begin{equation}
f_f = 
\frac{N_f}
{2\pi\Sigma_{f\alpha}\Sigma_{f\delta}}
\exp\left(-\frac{(\mu_{\alpha i}-\mu_{f\alpha})^2}{2\Sigma_{f\alpha}^2}-\frac{(\mu_{\delta i}-\mu_{f\delta})^2}{2\Sigma_{f\delta}^2}\right),
\label{eq:probdensf}
\end{equation}
with mean proper motion $\mu_{f\alpha}=2.85$ mas yr$^{-1}$, $\mu_{f\delta}=-8.86$ mas yr$^{-1}$ and dispersion $\Sigma_{f\alpha}=10.48$ mas yr$^{-1}$, $\Sigma_{f\delta}=10.78$ mas yr$^{-1}$. 
Since the intrinsic spread in the proper motions of the cluster
members is smaller than the measurement errors, the density function
for cluster members depends on the uncertainties in the proper-motion
components $\sigma_{tan i}$ and $\sigma_{r i}$ for each individual star :
\begin{equation}
f_c = 
\frac{N_c}
{2\pi\sigma_{ctan}\sigma_{cr}}
\exp\left(-\frac{(\mu_{tan i}-\mu_{ctan})^2}{2\sigma_{tan i}^2}-\frac{(\mu_{r i}-\mu_{cr})^2}{2\sigma_{r i}^2}\right).
\label{eq:probdensc}
\end{equation}
Following \citet{girard89clusmem}, the cluster membership probability for an individual star is 
$$
p=\frac{f_c}{f_c+f_f}.
$$

 Since $\sigma_{tan i}$ and $\sigma_{r i}$ are much smaller than
$\Sigma_{f\alpha}$ and  $\Sigma_{f\delta}$,  $f_c$ decreases
  much faster than $f_f$ and stars far from both distributions usually
  have $p$=0. Because the difference between the field and the cluster proper motion is
  much greater than their dispersions, there is however a locus in proper motion space where
  obvious outliers of both distributions could be selected as cluster
  members. We eliminated 2 such objects manually at the end of the
  selection process.
As in CC09, we extended this definition of cluster membership to
include information on the apparent magnitude.\\

\subsubsection{Apparent magnitude}
The same rationale was applied to determine whether the colour and apparent
magnitude of each source was closer to the field distribution
or to the colour magnitude relation expected for Hyades
stars.
For the Hyades this relation  was derived using the \citet{Pinsonneault.2004}
isochrone for [M/H]=+0.1 and age=600Myr that has been empirically
calibrated to fit the Hyades.
 We determined the expected apparent $K_s$ magnitude for Hyades
members of a given $V-K_s$ colour and compared it to the observed
magnitude of each object. Note that all near-infrared magnitudes used
in this article are 2MASS magnitudes. Hipparcos parallax information was
available for only 0.4\% of the overall number of sources, but eventually
for more than
30\% of the selected members (mainly because Hyades stars are much
closer than the average field stars and are more likely to have been
observed by Hipparcos). This was directly
converted into apparent $K_s$ magnitudes and compared to the observed
$K_s$ magnitude 
of the source. In most cases however, we do not have the parallax
information and we are forced to rely on the average Hyades distance,
associated with an arbitrary large -- and conservative -- dispersion (0.5
mag, assuming a 15-20pc cluster extent and ensuring we do not clip out
any good candidate with this approximate selection) of the expected
apparent magnitude to account for the large distance range where
Hyades members can be found. This integrates into the membership
estimation as follows: \\
We multiply the probability densities for the $K_s$ magnitude offset
from the field distributions with the field proper-motion probability
density from Eq.~\ref{eq:probdensf} above before obtaining the final
membership probability p:\\
\begin{equation}
g_f =\frac{f_f}{\sqrt{2\pi}\Sigma_{fK}}
\exp\left(-\frac{\delta K_{fi}^2}{\Sigma_{fK}^2}\right),
\end{equation}
and similarly for the cluster probability density we obtain
\begin{equation}
g_c =\frac{f_c}{\sqrt{2\pi}\Sigma_{cK}}\exp\left(-\frac{(K_i-K_{\rm isochrone})^2}{\Sigma_{cK}^2}\right).
\end{equation}
The combined membership probability for an individual star is  then 
$$
p=\frac{g_c}{g_c+g_f}.
$$

\begin{figure}
 \includegraphics[width=6.25cm,angle=270]{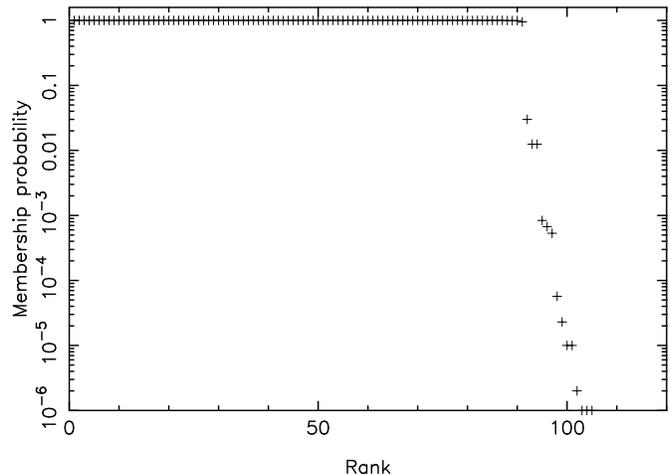}
 \caption{Ranked cluster membership probability for the 120 most probable
  Hyades cluster members variables sources.\label{pmem}}
\end{figure}

  Given the relatively large proper motion of the Hyades members and
  the large offset from the field
  population average proper 
  motion (over 100mas.yr$^{-1}$ difference from the field,
typically 10 times the proper motion measurement scatter, see figure
\ref{PM_Hyades}), the proper motion criterion is highly selective, while the
  large spatial extent (and hence depth) of the Hyades makes the apparent magnitude
  criterion much less accurate. We therefore used a very conservative
  magnitude selection parameter, with a large $\Sigma_{fK}$ of 0.5mag,
  to eliminate only the candidates far from the expected Hyades magnitudes
  and used proper motion as our main selection criterion. The
  resulting membership probability distribution is
largely bimodal, with objects clustering at probability of 1 (Hyades
members) or 0 (field sources), see figure \ref{pmem}.  We intended to
select all sources with p$>$50\% as cluster members, but in practical
there is no variable source in our Hyades sample with  94\%$>$p$>$50\%.

\begin{figure}
 \includegraphics[height=6.25cm,angle=0]{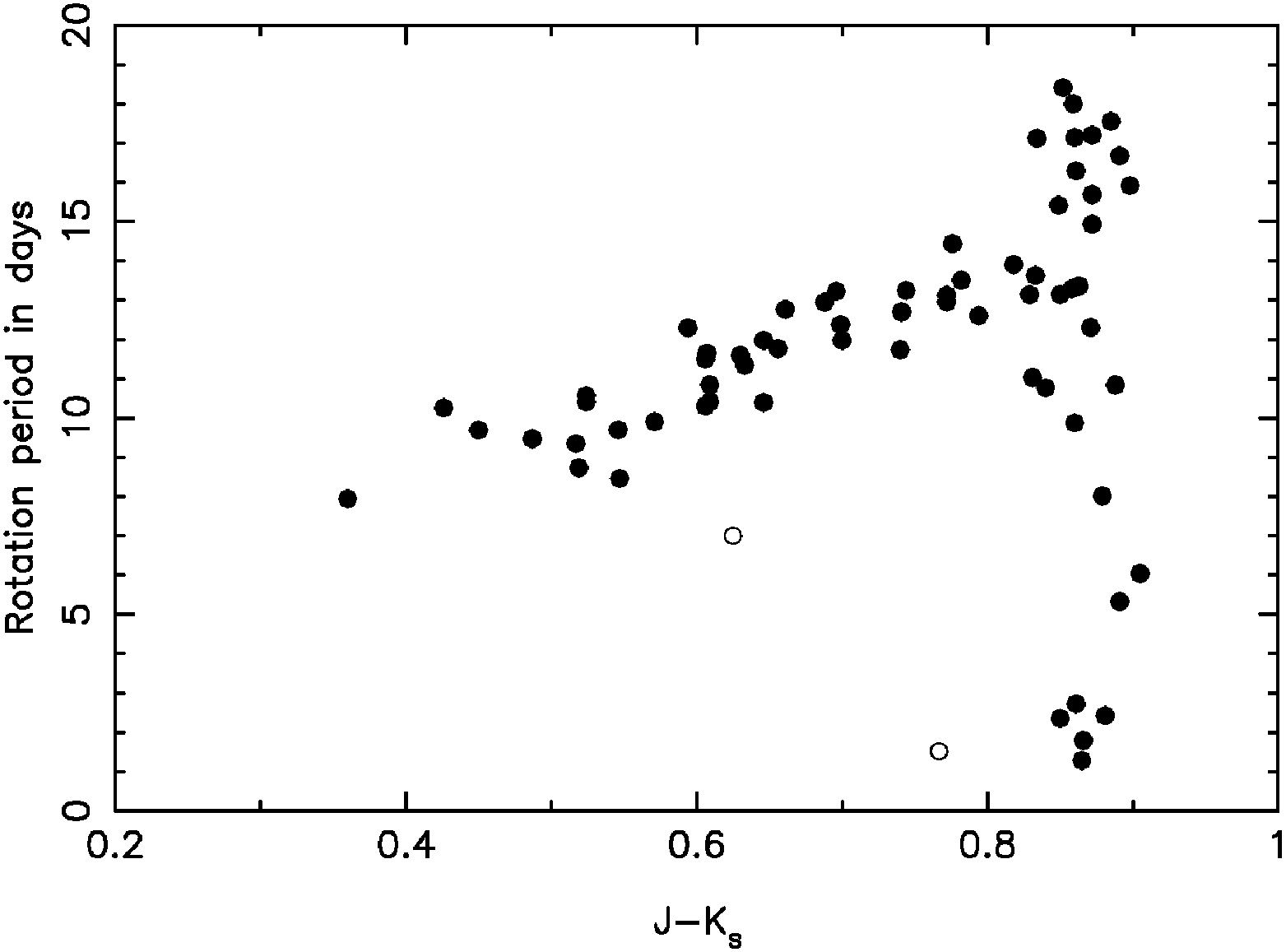} 
 \caption{$J-K_s$ colour-period plot of all selected Hyades
   members. The 2 outliers in hollow circles are known spectroscopic
   binaries. \label{col_perJK}}
\end{figure}

This automated selection yielded 91 variable sources identified as
cluster members.  This sample is far from being a complete census
of rotation periods of the hundreds of Hyades stars, and we put more
emphasis on clearing contamination than on increasing the sample
size. Since our FAP selection cut for the period
detection is 5\%, our sample is nevertheless expected to contain a few
spurious periods and more are expected because of possible
instrumental artefacts. We therefore visually
examined their light curves and periodograms, and computed their
autocorrelation functions (ACF) of the light curves using an inverse
variance-weighted adaptation of the Discrete Correlation Function
method of \citet{Edelson.1988}. We retained 83
cluster members whose light-curve autocorrelation functions
showed clear periodic structure, and whose periodogram-selected
periods agreed with the time lag of the first peak in the ACF. To
minimise any human bias during this step of the
analysis we did not check the objects' colour and did not know beforehand
whether or not they agreed with the colour-period relation described in next
section.
  As explained above, we also rejected two of our selected members
  because their proper motions, even though much closer to the Hyades 
proper motion than to the field, were still more than 10 $\sigma$ away
from the Hyades locus.
 Among the remaining 81 Hyades members with periods detected, 11 were
 independently detected twice (leaving 70 independent members), during two
different observing seasons. The
periods at the 2 epochs matched very 
well for 9 of them, with a scatter of the difference of 0.09day
(i.e. 0.8\% of their average period),
giving a good estimate of our period measurement accuracy. We assigned
the average period of the 2 observations to these stars. The 2
remaining objects (1SWASP J044735.33+145320.7 and 1SWASP
J051109.69+154857.5) were detected during one season at a higher frequency
harmonic (1/2 and 1/6 of the period) than the other season. The lower
frequency 
folded light curves appeared much more reliable than the harmonics and
were retained.
 We finally obtained a catalogue of 70 Hyades cluster
 members with reliable periods measured with SuperWASP. They are shown in
 Table \ref{Hyades_members}, Fig. \ref{col_perJK} and
 Fig. \ref{H_CMD}. Additional optical photometry and
 cross-identification with known cluster members is shown in the Appendix.

\begin{figure}
 \includegraphics[width=6.25cm,angle=270]{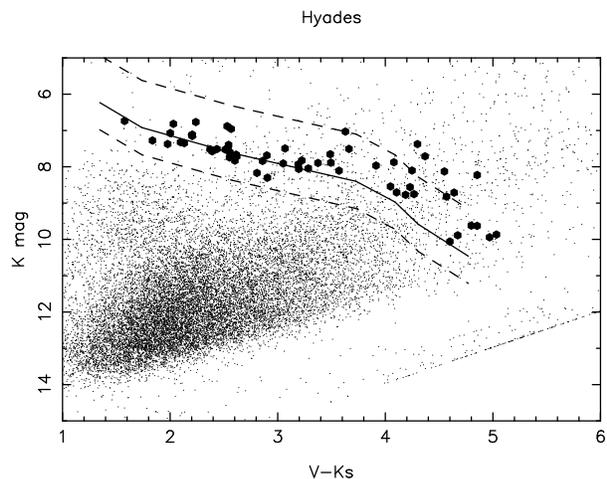}
 \caption{Colour-Magnitude diagram of field stars (dots) and Hyades selected
   members (black circles). The Hyades main sequence at cluster centre (42pc)
   is shown as a black line. The Hyades main sequences at 25pc and
   65pc are shown as dashed lines. \label{H_CMD}}
\end{figure} 

{\small
\begin{table*}\scriptsize
\centering
\begin{tabular}{|c|c|c|c|c|c|c|c|c|c|c|c|}\hline
Swasp id&USNO id&Xray&	Period&$J-K_s$&$K_s$&$V$&Dist.
&Proper Motion($\mu_{ra},\mu_{dec}$)&prlx&Proba.\\\hline
1SWASPJ033734.97+212035.4&1113-0041831&Yes&10.57&0.524&7.12&      	 9.3	&12.4&141.5$\pm$1.7  ,  -27.3$\pm$1.3&25.18&1.00\\
1SWASPJ034347.07+205136.4&1108-0041451&No&12.30&0.871&9.61&		  -    	&10.9&140.0$\pm$1.0  ,  -32.0$\pm$1.0&-&1.00\\
1SWASPJ035234.31+111538.6&1012-0033507&No&13.29&0.858&9.01&		   -   	&10.2&166.0$\pm$15.0  ,  -10.0$\pm$1.0&-&1.00\\
1SWASPJ035453.20+161856.3&1063-0039925&Yes&6.04$^{0.013}$&0.905&9.06&	    -  	&7.7&130.0$\pm$2.0  ,  -18.0$\pm$2.0&-&1.00\\
1SWASPJ035501.43+122908.1&1024-0045401&-&11.66&0.607&7.57&		 10.0	&8.9&127.5$\pm$2.1  ,  -9.6$\pm$1.5&25.62&1.00\\
1SWASPJ040339.03+192718.1&1094-0044958&No&11.45$^{0.075}$&0.606&7.60&	10.2	&6.0&119.0$\pm$2.3  ,   -34.1$\pm$1.8&23.27&1.00\\
1SWASPJ040525.67+192631.7&1094-0045433&No&13.51$^{0.016}$&0.782&8.11&	11.7	&5.6&123.0$\pm$3.2  ,   -28.4$\pm$2.2&15.76&1.00\\
1SWASPJ040634.62+133256.8&1035-0040066&No&16.68&0.891&9.09&		     - 	&6.0&112.0$\pm$2.0  ,     -10.0$\pm$2.0&-&1.00\\
1SWASPJ040701.22+152006.0&1053-0044486&No&11.98$^{0.036}$&0.700&7.66&	10.3	&5.1&122.6$\pm$2.0  ,   -18.9$\pm$1.5&24.19&1.00\\
1SWASPJ040743.19+163107.6&1065-0042472&No&12.30&0.594&7.51&		 9.9	&4.6&123.0$\pm$1.6  ,  -25.2$\pm$1.2&19.67&1.00\\
1SWASPJ040811.02+165223.3&1068-0041870&No&13.63&0.833&7.93&		11.1	&4.5&128.8$\pm$5.5  ,  -45.6$\pm$7.0&-&1.00\\
1SWASPJ040826.66+121130.6&1021-0043331&No&12.96&0.772&8.05&		11.3	&6.6&115.3$\pm$2.8  ,  -13.0$\pm$2.3&25.93&1.00\\
1SWASPJ040836.21+234607.0&1137-0048046&-&9.35&0.517&7.32&		 9.4	&7.9&120.4$\pm$0.6  ,  -43.9$\pm$0.7&-&1.00\\
1SWASPJ041127.64+155931.8&1059-0051230&No&1.79$^{0.008}$&0.866&9.94&	14.9	&3.9&120.0$\pm$4.0  ,  -16.0$\pm$3.0&-&1.00\\
1SWASPJ041510.33+142354.5&1043-0041060&No&13.91&0.818&8.13&		12.7	&3.9&118.8$\pm$4.6  ,  -11$\pm$.9$\pm$3.9&27.31&1.00\\
1SWASPJ041633.47+215426.8&1119-0053591&No&10.26&0.426&7.27&		 9.1	&5.4&105.6$\pm$1.3  ,  -37.5$\pm$1.1&19.07&1.00\\
1SWASPJ041725.14+190147.4&1090-0048172&No&12.95&0.688&7.91&		11.0	&2.9&112.0$\pm$6.3  ,  -27.8$\pm$8.1&-&1.00\\
1SWASPJ041728.13+145403.9&1049-0042422&Yes&2.35&0.850&9.62&		14.4	&3.1&104.0$\pm$6.0  ,  -18.0$\pm$2.0&-&1.00\\
1SWASPJ042322.85+193931.1&1096-0050945&Yes&9.90&0.571&7.15&		 9.4	&2.7&98.5$\pm$1.7  ,  -32.5$\pm$1.4&16.50&1.00\\
1SWASPJ042325.28+154547.2&1057-0062216&No&12.38&0.699&7.49&		10.6	&1.6&126.2$\pm$2.9  ,  -30.5$\pm$2.4&21.26&1.00\\
1SWASPJ042350.70+091219.5&992-0039564&Yes&5.33&0.891&8.23&		13.1	&7.9&104.0$\pm$3.0  ,  -6.0$\pm$1.0&-&1.00\\
1SWASPJ042359.13+164317.7&1067-0045253&No&17.14$^{0.320}$&0.860&8.56&	12.8	&0.8&110.0$\pm$2.0  ,  -26.0$\pm$1.0&-&1.00\\
1SWASPJ042416.93+180010.4&1080-0063868&No&11.60$^{0.039}$&0.630&7.52&	10.0	&1.1&112.7$\pm$2.4  ,  -33.2$\pm$1.9&19.31&1.00\\
1SWASPJ042500.18+165905.8&1069-0045244&No&11.77&0.656&7.83&		10.4	&0.5&95.8$\pm$2.8  ,  -23.7$\pm$2.6&-&1.00\\
1SWASPJ042514.54+185824.9&1089-0051192&Yes&10.84&0.888&8.70&		12.8	&1.9&92.0$\pm$4.0  ,  -28.02.0&-&1.00\\
1SWASPJ042547.55+180102.2&1080-0064462&Yes&8.46$^*$&0.547&6.77&	-	&0.9&112.6$\pm$2.4  ,  -35.7$\pm$2.0&20.79&1.00\\
1SWASPJ042642.81+124111.7&1026-0052108&Yes&12.61&0.794&7.03&		-	&4.4&123.4$\pm$3.9  ,  -16.7$\pm$2.5&29.09&1.00\\
1SWASPJ042648.25+105215.9&1008-0039886&Yes&10.40$^{0.075}$&0.646&6.88&	 9.4	&6.2&111.0$\pm$1.8  ,  -16.2$\pm$1.3&24.11&1.00\\
1SWASPJ042725.34+141538.3&1042-0043123&Yes&12.776$^*$&0.661&7.71&	10.3	&2.8&103.2$\pm$1.4  ,  -19.4$\pm$1.6&-&1.00\\
1SWASPJ042747.03+142503.8&1044-0042884&Yes&9.70&0.546&7.35&		 9.5	&2.7&101.4$\pm$1.9  ,  -20.0$\pm$1.5&19.00&1.00\\
1SWASPJ042828.78+174145.1&1076-0062237&Yes&2.42&0.881&7.71&		12.1	&0.7&105.1$\pm$16.3  ,  -20.1$\pm$8.3&-&1.00\\
1SWASPJ043033.88+144453.1&1047-0044652&No&18.41&0.852&9.71&		 -     	&2.5&102.0$\pm$12.0  ,  -22.0$\pm$2.0&-&1.00\\
1SWASPJ043034.87+154402.3&1057-0063658&Yes&8.73$^*$&0.519&6.82&	 8.8	&1.6&99.5$\pm$11.1  ,  -21.5$\pm$1.1&-&1.00\\
1SWASPJ043152.40+152958.3&1054-0052709&No&13.13&0.772&7.89&		11.3	&2.0&100.7$\pm$14.5  ,  -23.4$\pm$3.2&19.12&1.00\\
1SWASPJ043225.66+130647.6&1031-0059808&Yes&1.48$^*$&0.769&7.65&	11.1	&4.2&102.3$\pm$13.5  ,  -18.7$\pm$12.7&17.75&1.00
\\
1SWASPJ043323.75+235927.0&1139-0053890&Yes&17.55&0.885&8.03&		-	&7.1&94.0$\pm$12.0  ,  -50.0$\pm$1.0&-&1.00\\
1SWASPJ043327.00+130243.6&1030-0056878&No&16.29&0.861&8.82&		13.4	&4.3&108.0$\pm$13.0  ,  -18.0$\pm$2.0&-&1.00\\
1SWASPJ043337.18+210903.0&1111-0054738&No&12.69&0.741&7.69&		10.6	&4.4&109.2$\pm$12.2  ,  -45.3$\pm$1.7&26.74&1.00\\
1SWASPJ043411.14+113328.4&1015-0040520&Yes&11.03&0.831&8.06&		11.3	&5.8&100.7$\pm$12.2  ,  -17.5$\pm$1.5&-&1.00\\
1SWASPJ043548.51+131717.0&1032-0063919&Yes&13.36&0.863&9.87&		14.9	&4.4&92.0$\pm$12.0  ,  -16.0$\pm$1.0&-&1.00\\
1SWASPJ043605.26+154102.4&1056-0061021&No&9.47&0.487&7.37&		 9.3	&2.6&95.0$\pm$11.0  ,  -23.1$\pm$1.2&-&1.00\\
1SWASPJ043950.97+124342.5&1027-0058818&Yes&10.85&0.609&7.48&		10.0	&5.4&102.0$\pm$12.2  ,  -17.61.6&23.33&1.00\\
1SWASPJ044127.81+140434.1&1040-0045181&-&1.28&0.865&8.71&		13.3	&4.6&86.0$\pm$13.0  ,  -18.0$\pm$4.0&-&1.00\\
1SWASPJ044128.75+120033.7&1020-0044793&No&18.00&0.859&8.75&		13.0	&6.2&86.0$\pm$12.0  ,  -14.0$\pm$5.0&-&0.999\\
1SWASPJ044129.67+131316.3&1032-0065593&No&15.42$^*$&0.849&7.51&	11.2	&5.2&99.1$\pm$12.7  ,  -19.22.3&-&1.00\\
1SWASPJ044142.98+082620.0&984-0050517&-&8.02&0.879&10.06&		14.7	&9.4&100.0$\pm$14.0  ,  -6.0$\pm$2.0&-&1.00\\
1SWASPJ044315.69+170408.7&1070-0053038&Yes&10.31$^*$&0.606&7.40&	 9.9	&3.9&95.4$\pm$11.6  ,  -30.4$\pm$1.3&-&1.00\\
1SWASPJ044618.79+033810.7&936-0059947&No&13.25&0.744&7.83&		11.0	&14.3&92.5$\pm$12.8  ,  12.9$\pm$2.1&23.05&1.00\\
1SWASPJ044630.38+152819.3&1054-0056655&Yes&7.95&0.360&6.74&		 8.3	&5.0&91.3$\pm$11.4  ,  -24.7$\pm$1.0&21.08&1.00\\
1SWASPJ044718.51+062711.6&964-0047473&No&14.44&0.776&7.87&		12.0	&11.7&99.9$\pm$15.0  ,  9.5$\pm$2.8&-&1.00\\
1SWASPJ044800.88+170321.6&1070-0054253&Yes&10.77&0.840&7.37&		11.7	&5.0&89.1$\pm$12.1  ,  -30.22.1&-&1.00\\
1SWASPJ044830.61+162319.0&1063-0051391&-&15.69&0.872&8.54&		12.6	&5.2&78.0$\pm$16.0  ,  -30.0$\pm$3.0&-&1.00\\
1SWASPJ044842.13+210603.6&1111-0060275&Yes&9.69&0.450&7.07&		 9.1	&6.5&81.8$\pm$15.0  ,  -46.4$\pm$4.7&19.41&1.00\\
1SWASPJ044912.98+244810.2&1148-0059776&Yes&6.9$^*$&0.626&6.96&		 9.5	&9.3&85.2$\pm$11.6  ,   -52.0$\pm$11.2&20.11&1.00\\
1SWASPJ044952.11+060633.6&961-0047586&No&17.12&0.834&9.63&		14.5	&12.3&96.0$\pm$15.0  ,  6.0$\pm$2.0&-&1.00\\
1SWASPJ045000.70+162443.5&1064-0050763&No&11.98&0.646&7.84&		10.7	&5.5&89.2$\pm$12.2  ,  -28.9$\pm$1.7&-&1.00\\
1SWASPJ045102.41+145816.5&1049-0049205&-&13.14$^{0.102}$&0.829&7.97&	11.9	&6.2&88.0$\pm$13.0  ,  -32.0$\pm$6.0&-&1.00\\
1SWASPJ045223.53+185948.9&1089-0058167&Yes&11.34&0.633&7.74&		10.3	&6.3&78.6$\pm$12.3  ,  -31.4$\pm$1.5&21.55&1.00\\
1SWASPJ045223.86+104309.9&1007-0045668&-&9.88&0.860&8.77&		13.0	&8.9&76.0$\pm$14.0  ,  -12.0$\pm$5.0&-&1.00\\
1SWASPJ050540.37+062754.6&964-0053308&No&10.41&0.609&7.52&		 9.9	&14.2&66.0$\pm$12.1  ,  4.2$\pm$1.1&19.24&1.00\\
1SWASPJ051109.69+154857.5&1058-0068980&-&14.94&0.872&8.10&		12.3	&10.7&63.2$\pm$16.3  ,  -26.8$\pm$18.3&-&0.992\\
1SWASPJ051119.30+075432.0&979-0066501&-&13.23&0.696&8.30&		11.2	&14.2&52.5$\pm$16.3  ,  0.2$\pm$18.1&-&0.947\\\hline
\end{tabular}
 \caption{Hyades members with SuperWASP rotation periods. Period is in
   days and if the object has been identified on 2 different seasons,
   the period given is the average of the  2 periods whose difference
  is indicated in exponent. Proper motion and parallax are in mas. Dist. is the
   distance to the cluster center in degrees, and Proba. is the
   membership probability. $J-K_s$ is a 2MASS
   colour. X ray traces whether or not \citet{Stern.1995}
   reported X ray activity.   \label{Hyades_members}}
\end{table*}
}

\section[]{Period-colour relation}
\subsection{The relation}
\begin{figure}
 \includegraphics[width=6.25cm,angle=270]{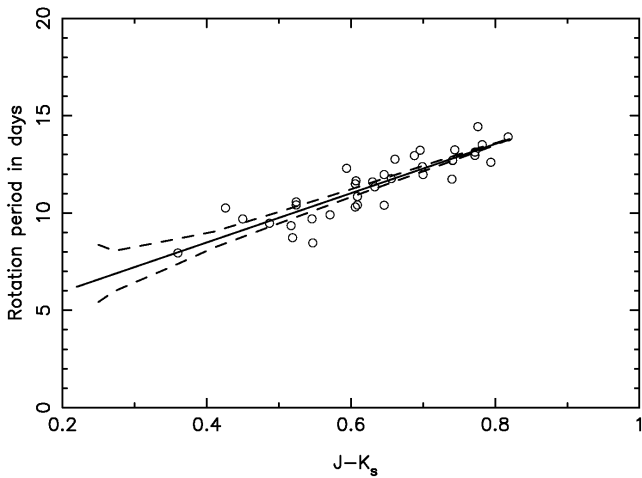} 
 \caption{$J-K_s$ colour-period plot of Hyades members that are used
   to derive the colour-period relation (black line). The dashed
   lines represent the expected spread in periods caused by
   differential rotation (see CC09 for details). \label{col_per_rel_H}}
\end{figure}

 To derive a clean period-colour relation, we selected the reliable
 candidates which sit in the $0.2<J-K_s<0.82$ colour range, where Hyades stars have converged toward an
  identifiable colour-period relation.
Since we have a large sample we used a
 recursive clipped average selection to eliminate those objects that
 are likely outliers (binary systems, mistaken cluster members,
 or instrumental period artefacts).  A least square
 linear fit was first performed on all sources in this colour
 range. Their dispersion around this first estimate of the relation
 was calculated and all candidates more than 2$\sigma$ away from this
  relation were removed. It turns out that this only removed
    the two 
    members identified as spectroscopic binaries  that we could have
    removed manually. Since we also used the
    same method, with the same parameters, to remove outliers on our
    Praesepe sample (see next section ), for which information  
  on spectroscopic binarity was not available, this allowed a
  consistent analysis of both clusters samples. The 2 Hyades outliers
  that are spectroscopic binaries are 1SWASPJ043225.66+130647.6
  \citep[HD286839, orbital period of 1.48day, ][]{Mermilliod.2009}
 and
  1SWASPJ044912.98+244810.2 \citep[HD283882, SB2 with an orbital period of
  11.93 days ][]{Mermilliod.2009}.
 A new least square
  linear fit was 
  performed on the remaining sources to derive the final relation, with a smaller
  dispersion (see Fig.\ref{col_per_rel_H}).
 We also tried to fit a quadratic relation through our
  data, but it did not significantly increase the correlation
  coefficient of the fit. We therefore used the simpler linear fit to
  find the period-colour relation.

The derived relation is the following:
\begin{equation}
P =   11.401 +   12.652 * (J-K_s-0.631)
\label{P}
\end{equation}
 Though the  0.58day dispersion of periods around the relation is
  smaller than the dispersion observed around more
  complex colour-period relation in younger clusters such as M34, M35
  and M37 \citep{Meibom.2009_M35,Hartman.2009}, it is
significantly higher than the 0.19day dispersion observed around Coma
cluster relation by CC09 using the same methods. As explained above, we
used objects whose periods
have been detected during 2 different seasons to estimate
that the error on the period measurement is about 0.1 day. This means
that the relatively large spread of periods for a given colour that we
observe in the Hyades is not an observational bias but is
real. It could not be caused by differential rotation because this
affects significantly only F and G stars
\citep[][]{Barnes.2005,Reiners.2006} while the spread in periods 
is observed on the FGKM spectral range (see
Fig.\ref{col_per_rel_H}). Theoretical models of stellar
spin-down \citep[see][for instance]{Kawaler.1988,Cameron.1994,Irwin.2009} do predict that by 600Myr stellar spin has almost, but not
completely, converged for stars more massive than than
0.25$M\odot$. However, this does not explain the difference between
Coma and the Hyades, which is 
discussed more extensively in section 5., and includes data
on Praesepe cluster.


\subsection{Comparison to Radick 87\&95 members}
\begin{figure*}
\begin{tabular}{cc}
 \includegraphics[width=6.25cm,angle=270]{col_perJK_Hradick.ps} &
\includegraphics[width=6.25cm,angle=270]{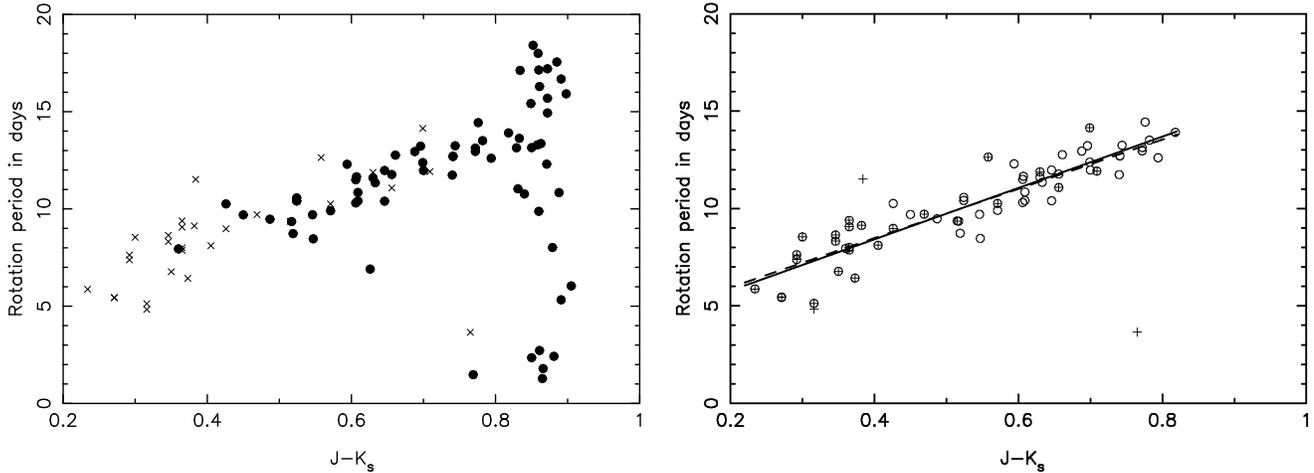}
\end{tabular}
 \caption{{\bf Left:}$J-K_s$ colour-period plot of all SuperWASP-selected Hyades
   members (black dots) and rotational periods previously observed by
   \citet{Radick.1987,Radick.1995} (crosses). {\bf Right:} Stars used
   to derive the linear fit with SuperWASP and R87-95 data (black line)
   are circled. The SuperWASP-only relation is shown in dashed
   line.\label{col_perJK_radick}} 
\end{figure*}

 Fig. \ref{col_perJK_radick} shows our new SuperWASP periods for
 Hyades cluster members together with previous objects measured by
 \citet[][hereafter R87-95]{Radick.1987,Radick.1995}. The scatter of
 the R87-95 periods appears larger, perhaps because of differential
 rotation on these higher mass objects or perhaps because the time
 sampling of R87-95, optimised toward monitoring years-long
 photometric variability and not day-long rotational periods causes
 larger errors on period measurements. Direct comparison with our 
 data is, unfortunately, difficult since most
 of the R87-95 are blue, bright stars that are saturated in SuperWASP images.
We fitted a linear colour-period relation through the full
SuperWASP+R87-95 data and obtained a quite similar relation:\\
\begin{equation}
P' =   10.189 +   13.091 * (J-K_s-0.536)
\label{P'}
\end{equation}
with a dispersion of 0.79 day around the relation, which is shown,
together with SuperWASP-only data in Fig. \ref{col_perJK_radick}.

 We also tried to fit a quadratic relation,
 but this did not improve the correlation. By eye it seems there could be a
 break in the colour-period relation around $J-K_s$=0.43. We therefore
 tried to fit a two-part piecewise linear relation through the data, but again this did
 not improve the correlation coefficient: the apparent break in the relation is
 likely caused by the increased scatter of R87-95 data. We
 consequently used the simple linear fit. However, given the small
 difference between the relations in Eqs.~\ref{P} and \ref{P'}, in the following we
 use Eq.~\ref{P}, from SuperWASP-only data, to compare with SuperWASP-only data
 from the Praesepe and Coma clusters.

\section{Praesepe }
\subsection{Observations, reduction and candidate selection}
   Our analysis of the SuperWASP fields overlapping Praesepe Cluster
   (also known as M44 or Melotte 88) followed the same strategy as our
   analysis of the Hyades that has been described in the previous
   section. Table \ref{P_fields} details the  2 SuperWASP fields with
   a good time sampling (60-70 images per night) and a long enough
   baseline (130 nights) that 
   we found within 10 degrees of the cluster center (08$^{\rm h}$ 40$^{\rm m}$.4 +19$^\circ$ 41$'$).
    We looked for all rotational variables
   with periods  between 1.1days and 20 days in these fields using the
   light curve analysis described in section \ref{light_curve}. This
   yielded 3324 rotational variables. 

\subsubsection{Candidate selection.}
\begin{figure}
 \includegraphics[width=6.25cm,angle=270]{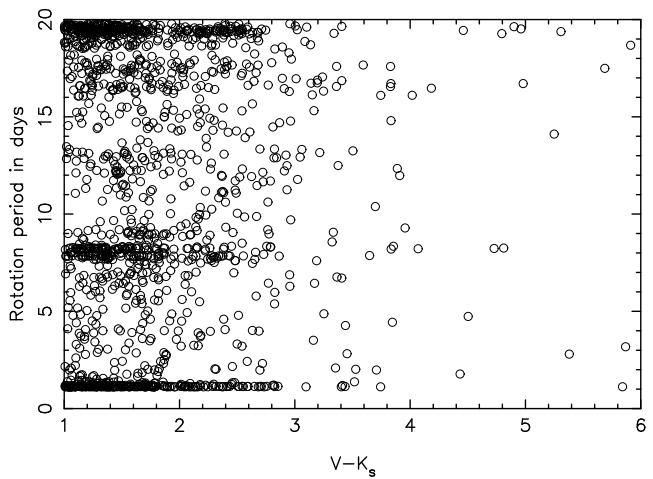}
 \caption{All sources with a good signal to noise rotational signal
   detected in the fields around Praesepe cluster.\label{P_initial_sample}}
\end{figure}

As can be seen in
   Fig. \ref{P_initial_sample} there is a significant bias in these
   data that causes
   over-numerous detections of periods greater than 18 days. 
   A close
   examination of these light curves confirmed that this was likely
   due to an instrumental/observational bias on one of the detectors. We
   consequently eliminated all these sources for the rest of the
   study.  The
     over-densities visible on Fig. \ref{P_initial_sample} around 8 and
     13 days are likely due to windowing in 
   SWASP time-sampling, but only one candidate, was selected at a
   nearby period, 1SWASP J083627.86+210716.2. This period however appears
   reliable since it has been independently detected on the 2 fields.
   Following again the same procedure as we used for the
   Hyades, we selected candidates closer to the colour and proper motion of
   Praesepe \citep[$\mu_\alpha=-35.99\pm 4$ mas.yr$^{-1}$;
     $\mu_\delta=-12.92\pm 4$
     mas.yr$^{-1}$][]{Loktin.2003,Kraus.2007}. Since Praesepe is not
   as
   nearby nor as extended on the sky as the Hyades we performed a
   simple comparison of RA/DEC proper motion and not a comparison of
   proper motion tangential and perpendicular to the convergent point
   as we did for the Hyades. The overall membership probability
   determination from proper motion and apparent magnitude is
   otherwise identical to what we did for the Hyades and described in
   section 3. The expected colour-magnitude relation for clusters members was
derived using the isochrones of \citet{Pinsonneault.2004}, for an age of
600Myr and a metallicity of [M/H]=+0.1 (+0.14 for Praesepe according to
\citet{Scholz.2007}).

\begin{figure}
 \includegraphics[width=6.25cm,angle=270]{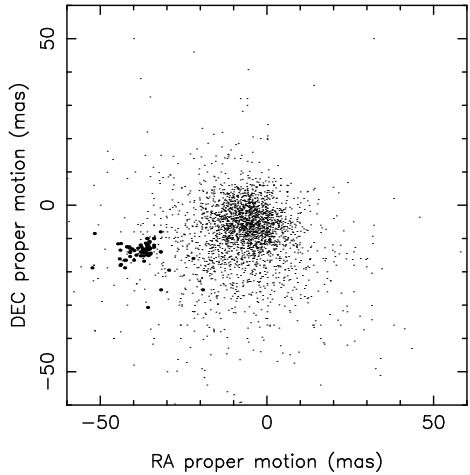}
 \caption{Proper motion in RA and DEC of all sources with identified rotational
   modulation signal in our survey around Praesepe. Large dots are objects identified as cluster member owing to their proper motion and apparent magnitude.\label{PM_Praesepe}}
\end{figure}

\begin{figure}
 \includegraphics[width=6.25cm,angle=270]{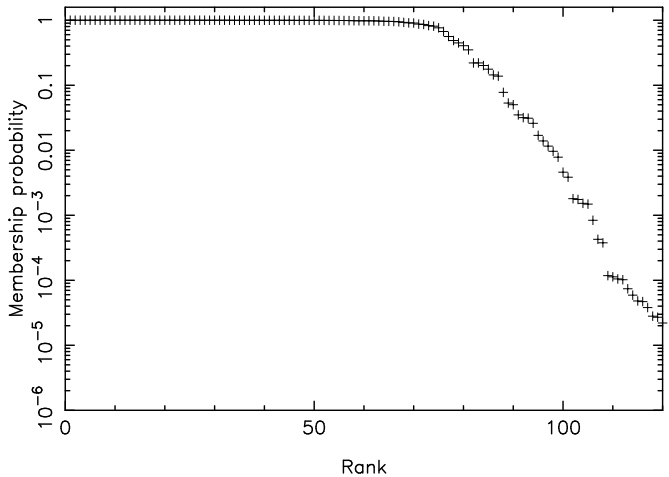}
 \caption{Ranked cluster membership probability for the 120 most probable
  Praesepe cluster member variables sources.
   \label{P_pmem}}
\end{figure}

\begin{figure}
 \includegraphics[width=6.25cm,angle=270]{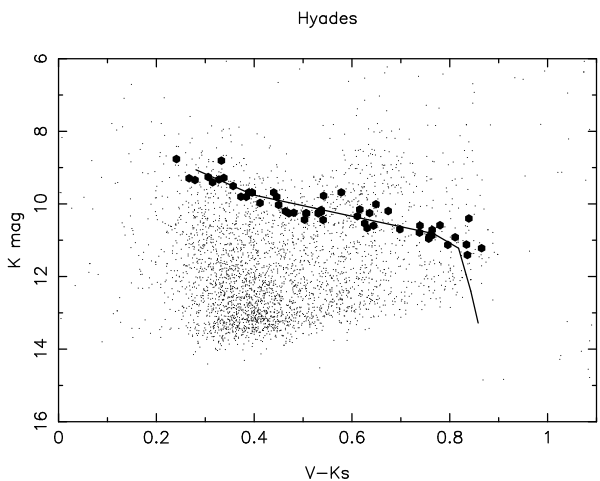}
 \caption{Colour-Magnitude diagram of field stars (dots) and Praesepe selected
   members (black circles). The  Praesepe main sequence is shown as a
   black line.  \label{P_CMD}}
\end{figure}

 This analysis yielded 71 reliable Praesepe cluster members with
a membership probability over 50\%. These objects are shown
in Fig.\ref{PM_Praesepe}  and additional optical photometry and
 cross-identification with known cluster members is shown in the
 Appendix. Since the proper motion locus of Praesepe is 
closer to the field stars' locus, the membership probability
distribution shown on Fig. \ref{P_pmem} is not as bimodal as the
distribution obtained for the Hyades but still non-ambiguously
distinguishes between cluster members and others.  One of this
  selected members had $V-K_s$ colours and a $K_s$ magnitude
  compatible with an early M-dwarf cluster member but $J-K_s$
  colours  and magnitudes typical of a field G dwarf. Since $J-K_s$
  colour is more reliable, 
this object was rejected as a probable field star. Overall Praesepe
data was of lower 
quality than the average Hyades data, first because of a slightly
sparser time sampling and secondly because of the persistence of some
moon-related red noise even after decorrelation of data.  A careful analysis
of each light curve eliminated 12 of the candidates whose derived 
periods appeared spurious. Again this selection was done
without the knowledge of the colour to avoid any human bias toward
removing preferentially candidates that would not fit the relation.
Six objects turned out to have been detected twice, at about the same
period in both field. As for the Hyades we assigned the average of the 2
periods to these objects. We use the scatter of the difference of both
periods, 0.14day, as an estimate of our error on period measurement.
Our Praesepe analysis uses the 52 independent remaining cluster members
with reliable periods visible on table \ref{P_members} and Fig. \ref{P_CMD}.

\begin{table}
 \centering
 \caption{Praesepe superWASP fields used in this survey.\label{P_fields}}
\begin{tabular}{|c|c|c|c|c|c|c|} \hline
Field centre & No. of & Usable & First  & Baseline\\ 
             & images & nights & night & (nights) \\ \hline
08 42' +24 18' &3963  & 59 &2006-11-26 & 130 \\ \hline
08 52' +17 35' &4144  & 60 &2006-11-26 & 130 \\ \hline
\end{tabular}
\end{table}

\begin{table*}
\centering
\begin{tabular}{|c|c|c|c|c|c|c|c|c|c|c|} \hline
Swaspid & USNO id   &	Period & $J-K_s$ & $K_s$ & $V$ & Dist.&Proper Motion($\mu_{ra},\mu_{dec}$)  & Proba. \\ \hline
1SWASP J083141.93+260641.0 &  1161-0156334 &  13.16 &   0.738 &  10.79 &		 13.7&   6.7  &  -42.8$\pm$ 5.6 ,  -16.5$\pm$ 5.6 &  1.000 \\
1SWASP J083554.98+224611.3 &  1127-0193591 &   9.31 &   0.532 &  10.23 &		 12.7&   3.3 &  -29.5$\pm$ 5.2  ,  -19.5$\pm$ 5.2 &  0.928 \\
1SWASP J083622.69+191129.3 &  1091-0164534 &   9.83 &   0.506 &  10.25 &		 12.4&   1.1 &  -39.9$\pm$ 6.0  ,  -13.6$\pm$ 6.1 &  0.998 \\
1SWASP J083627.81+175453.4 &  1079-0203318 &   6.91 &   0.315 &   9.40 &		 10.9&   2.0 &  -36.5$\pm$ 1.1  ,  -11.4$\pm$ 1.1 &  0.999 \\
1SWASP J083627.86+210716.2 &  1111-0173917 &  13.09$^{0.123}$ &   0.757 &  10.96 &	 13.8&   1.7 &  -36.0$\pm$ 1.0  ,  -10.0$\pm$ 1.0 &  0.979 \\
1SWASP J083628.30+201342.8 &  1102-0162570 &  10.25 &   0.541 &  10.44 &		 12.7&   1.1 &  -44.3$\pm$ 6.3  ,  -16.1$\pm$ 6.3 &  0.998 \\
1SWASP J083648.96+191526.4 &  1092-0164565 &   7.41 &   0.396 &   9.68 &		 11.3&   0.9 &  -36.2$\pm$ 2.3  ,  -12.8$\pm$ 1.3 &  1.000 \\
1SWASP J083657.82+213355.9 &  1115-0175522 &  13.27$^{0.111}$ &   0.834 &  11.12 &	 14.1&   2.0 &  -42.1$\pm$ 6.3  ,  -12.5$\pm$ 6.3 &  1.000 \\
1SWASP J083711.48+194813.2 &  1098-0164221 &   8.84 &   0.389 &   9.68 &		 11.1&   0.8 &  -35.4$\pm$ 0.7  ,  -12.8$\pm$ 0.7 &  0.977 \\
1SWASP J083718.29+194156.3 &  1096-0163761 &   8.90 &   0.373 &   9.80 &		 11.6&   0.7 &  -37.1$\pm$ 1.9  ,  -15.2$\pm$ 1.9 &  0.999 \\
1SWASP J083722.23+201037.0 &  1101-0161763 &   3.91 &   0.384 &   9.80 &		 11.3&   0.9 &  -36.0$\pm$ 0.7  ,  -14.5$\pm$ 0.9 &  0.990 \\
1SWASP J083727.54+193703.1 &  1096-0163804 &   8.68 &   0.446 &   9.80 &		 11.5&   0.7 &  -34.0$\pm$ 1.9  ,  -12.6$\pm$ 1.9 &  0.998 \\
1SWASP J083733.07+183915.5 &  1086-0168273 &   7.59 &   0.338 &   9.28 &		 10.5&   1.2 &  -37.5$\pm$ 1.2  ,  -14.3$\pm$ 1.4 &  0.999 \\
1SWASP J083735.77+205927.5 &  1109-0174003 &  10.09$^{0.428}$ &   0.636 &  10.25 &	 12.9&   1.5 &  -32.0$\pm$ 4.0  ,  -14.0$\pm$ 3.0 &  0.994 \\
1SWASP J083746.41+193557.5 &  1095-0163684 &   9.50 &   0.481 &  10.24 &		 12.3&   0.6  &  -41.3$\pm$ 6.3  ,  -12.9$\pm$ 6.3 &  0.996 \\
1SWASP J083747.39+190624.8 &  1091-0164888 &   9.30 &   0.464 &  10.20 &		 11.9&   0.8 &  -35.6$\pm$ 2.0  ,  -15.1$\pm$ 2.0 &  0.911 \\
1SWASP J083857.23+201053.6 &  1101-0162189 &  11.78 &   0.764 &  10.71 &		 13.7&   0.6 &  -41.2$\pm$ 6.4  ,  -15.4$\pm$ 6.4 &  1.000 \\
1SWASP J083902.28+191934.4 &  1093-0163918 &   9.42 &   0.471 &  10.25 &		 12.3&   0.5 &  -44.1$\pm$ 6.3  ,  -13.5$\pm$ 6.3 &  0.998 \\
1SWASP J083921.54+204529.4 &  1107-0171494 &   9.11 &   0.450 &  10.02 &		 11.6&   1.1 &  -35.0$\pm$ 1.3  ,  -14.4$\pm$ 1.3 &  0.826 \\
1SWASP J083930.44+200408.6 &  1100-0163760 &   7.02 &   0.333 &   8.80 &		 10.2&   0.4 &  -35.7$\pm$ 0.8  ,  -13.4$\pm$ 0.7 &  0.992 \\
1SWASP J083935.53+185236.8 &  1088-0166993 &   6.29 &   0.328 &   9.32 &		 10.6&   0.8 &  -37.3$\pm$ 1.2  ,  -13.1$\pm$ 1.5 &  0.999 \\
1SWASP J083945.78+192201.1 &  1093-0164074 &   5.90 &   0.306 &   9.25 &		 10.8&   0.4 &  -35.3$\pm$ 0.9  ,  -12.8$\pm$ 0.9 &  1.000 \\
1SWASP J084005.72+190130.7 &  1090-0167104 &   4.80 &   0.649 &  10.00 &		 12.5&   0.7 &  -38.0$\pm$ 6.3  ,  -15.2$\pm$ 6.4 &  0.960 \\
1SWASP J084022.41+203827.1 &  1106-0171109 &  14.59 &   0.865 &  11.22 &		 14.3&   1.0 &  -38.0$\pm$ 1.0  ,  -14.0$\pm$ 2.0 &  1.000 \\
1SWASP J084025.56+192832.8 &  1094-0163661 &  13.15 &   0.241 &   8.76 &		  9.7&   0.2 &  -36.7$\pm$ 0.8  ,  -13.3$\pm$ 0.7 &  0.999 \\
1SWASP J084033.46+193801.0 &  1096-0164636 &   9.34 &   0.537 &  10.16 &		 12.3&   0.1 &  -41.1$\pm$ 6.3  ,  -13.9$\pm$ 6.4 &  0.999 \\
1SWASP J084036.23+213342.1 &  1115-0176177 &   8.88$^{0.119}$ &   0.421 &   9.97 &	 11.8&   1.9 &  -36.8$\pm$ 1.1  ,  -14.2$\pm$ 0.6 &  0.995 \\
1SWASP J084044.25+202818.7 &  1104-0163841 &   4.27 &   0.839 &  10.40 &		 13.8&   0.8 &  -36.0$\pm$ 4.0  ,  -10.0$\pm$ 4.0 &  0.906 \\
1SWASP J084047.61+185411.9 &  1089-0167723 &   8.89 &   0.440 &   9.68 &		 11.2&   0.8 &  -36.2$\pm$ 1.2  ,  -14.9$\pm$ 1.2 &  0.996 \\
1SWASP J084048.34+195518.9 &  1099-0165066 &   7.43 &   0.357 &   9.50 &		 11.0&   0.3 &  -35.5$\pm$ 1.2  ,  -13.0$\pm$ 0.6 &  1.000 \\
1SWASP J084055.31+183459.0 &  1085-0167929 &  12.49 &   0.796 &  11.12 &		 14.0&   1.1 &  -41.2$\pm$ 6.4  ,  -14.2$\pm$ 6.3 &  0.998 \\
1SWASP J084059.70+182204.5 &  1083-0186951 &   9.13 &   0.578 &   9.68 &		 11.5&   1.3 &  -41.4$\pm$ 5.2  ,  -12.5$\pm$ 5.7 &  0.995 \\
1SWASP J084122.57+185602.0 &  1089-0167811 &  11.16 &   0.626 &  10.53 &		 12.9&   0.8 &  -35.4$\pm$ 6.3  ,  -11.2$\pm$ 6.4 &  0.994 \\
1SWASP J084130.69+185218.6 &  1088-0167362 &   2.43 &   0.616 &  10.15 &		 12.7&   0.9 &  -36.0$\pm$ 2.0  ,  -12.0$\pm$ 3.0 &  0.999 \\
1SWASP J084143.67+195743.8 &  1099-0165295 &   9.15 &   0.507 &  10.25 &		 12.4&   0.4 &  -40.5$\pm$ 1.9  ,  -13.1$\pm$ 1.9 &  0.989 \\
1SWASP J084158.83+200627.2 &  1101-0162974 &  11.08 &   0.644 &  10.59 &		 13.0&   0.6 &  -44.1$\pm$ 6.3  ,  -18.1$\pm$ 6.3 &  0.999 \\
1SWASP J084220.09+190905.7 &  1091-0165816 &  12.04 &   0.698 &  10.69 &		 13.2&   0.7 &  -37.0$\pm$ 6.3  ,  -16.8$\pm$ 6.5 &  0.998 \\
1SWASP J084237.01+200832.0 &  1101-0163099 &  12.22 &   0.763 &  10.87 &		 13.7&   0.7 &  -42.4$\pm$ 6.4  ,  -16.5$\pm$ 6.4 &  1.000 \\
1SWASP J084245.96+211616.3 &  1112-0175801 &   3.52$^{0.001}$ &   0.811 &  10.92 &	 14.0&   1.7 &  -36.0$\pm$ 3.0  ,  -12.0$\pm$ 1.0 &  1.000 \\
1SWASP J084248.47+203424.5 &  1105-0167705 &   9.95 &   0.531 &  10.26 &		 12.4&   1.1 &  -35.2$\pm$ 2.0  ,  -13.1$\pm$ 2.0 &  1.000 \\
1SWASP J084308.22+194247.7 &  1097-0164856 &  11.71 &   0.631 &  10.66 &		 13.1&   0.6 &  -43.9$\pm$ 6.3  ,  -17.8$\pm$ 6.3 &  0.998 \\
1SWASP J084317.83+203037.3 &  1105-0167843 &   8.91 &   0.542 &   9.77 &		 12.0&   1.1 &  -37.3$\pm$ 1.3  ,  -14.2$\pm$ 2.0 &  0.981 \\
1SWASP J084332.39+194437.8 &  1097-0164949 &   9.75 &   0.537 &  10.21 &		 12.3&   0.7 &  -40.0$\pm$ 2.0  ,  -16.5$\pm$ 2.0 &  0.961 \\
1SWASP J084344.72+211234.5 &  1112-0175989 &   2.83$^{0.004}$ &   0.836 &  11.41 &	 14.8&   1.7 &  -38.0$\pm$ 2.0  ,  -12.0$\pm$ 3.0 &  1.000 \\
1SWASP J084519.17+190010.9 &  1090-0168262 &  12.30 &   0.758 &  10.89 &		 13.5&   1.3 &  -43.9$\pm$ 6.3  ,  -11.5$\pm$ 6.3 &  0.997 \\
1SWASP J084624.31+170235.1 &  1070-0184311 &  11.43 &   0.674 &  10.19 &		 12.9&   3.0 &  -51.8$\pm$ 6.0  ,   -8.5$\pm$ 6.0 &  0.945 \\
1SWASP J084647.32+193840.9 &  1096-0165969 &   6.28 &   0.279 &   9.34 &		 10.9&   1.5 &  -36.5$\pm$ 0.6  ,  -13.9$\pm$ 1.0 &  1.000 \\
1SWASP J084714.12+162347.3 &  1063-0162881 &   5.70 &   0.267 &   9.29 &		 10.6&   3.7 &  -37.7$\pm$ 1.2  ,  -12.6$\pm$ 1.3 &  1.000 \\
1SWASP J084817.25+165447.7 &  1069-0177337 &  11.25 &   0.739 &  10.59 &		 13.5&   3.3 &  -42.7$\pm$ 6.0  ,  -18.8$\pm$ 6.1 &  1.000 \\
1SWASP J084847.83+274136.5 &  1176-0212100 &  12.36 &   0.780 &  10.58 &		 13.7&   8.2 &  -52.5$\pm$ 5.2  ,  -18.8$\pm$ 5.2 &  0.994 \\
1SWASP J084948.35+160750.7 &  1061-0165009 &  11.38 &   0.503 &  10.43 &		 12.6&   4.2 &  -44.8$\pm$ 6.0  ,  -11.6$\pm$ 6.0 &  0.997 \\
1SWASP J090222.37+182223.9 &  1083-0190841 &  14.82 &   0.611 &  10.34 &		12.6&   5.4 &  -32.0$\pm$ 2.0  ,   -8.0$\pm$ 3.0 &  0.671 \\ \hline
\end{tabular}
 \caption{Praesepe members with SWASP rotation periods. Period is in
   days and if the object has been identified on 2 different seasons,
   the period given is the average of the  2 periods whose difference
  is indicated in exponent. Proper motion and parallax are in mas. Dist. is the
   distance to the cluster centre in degrees, and Proba. is the
   membership probability. $J-K_s$ is a 2MASS
   colour.\label{P_members}}
\end{table*}

\subsection{Period-colour relation}
  As for the Hyades we tried to fit a linear relation, a quadratic law
  and 2 linear laws (for $J-K_s<$0.43 and 
$J-K_s>$0.43), and the resulting correlation coefficients were very
  similar. We therefore favoured the simpler linear fit because it has
  fewer free parameters and fits well to the data.

  We derived the colour-period relation is a similar way as for the
  Hyades, using all members with 0.2$<J-K_s<$0.82, and making a
  linear fit after clipping-out the 5  outliers above and below the
  relation in this colour range. Note that 2 of them, 1SWASP
  J083722.23+201037.0, and 1SWASP J084005.72+190130.7, have been
  identified at a period that is half of the expected period. It is
  likely that these objects rotation periods actually sit on the
  period colour relation but that the period analysis found the
  half-period harmonic. This phenomena, already observed in the Hyades
  for 1SWASP J044735.33+145320.7
   and in Coma (see CC09), is relatively common and is caused by
   diametrally symetric stellar spot patterns.  Among the other outliers
   1SWASP
   J090222.37+182223.9 has a relatively low membership probability
   (67\%) and 1SWASP J084025.56+192832.8 stands out in the
   colour-magnitude 
   diagram (Fig. \ref{H_CMD})as a likely binary, which can affect its period or
   colour. The fast rotator 1SWASP J084130.69+185218.6 with
   $J-Ks=$0.616 seems a regular object and is difficult to explain. It
   could however be a contrasted tidally locked binary, but we have no
   way to ascertain this for now.

\begin{figure*}
\begin{tabular}{cc}
 \includegraphics[width=6.25cm,angle=270]{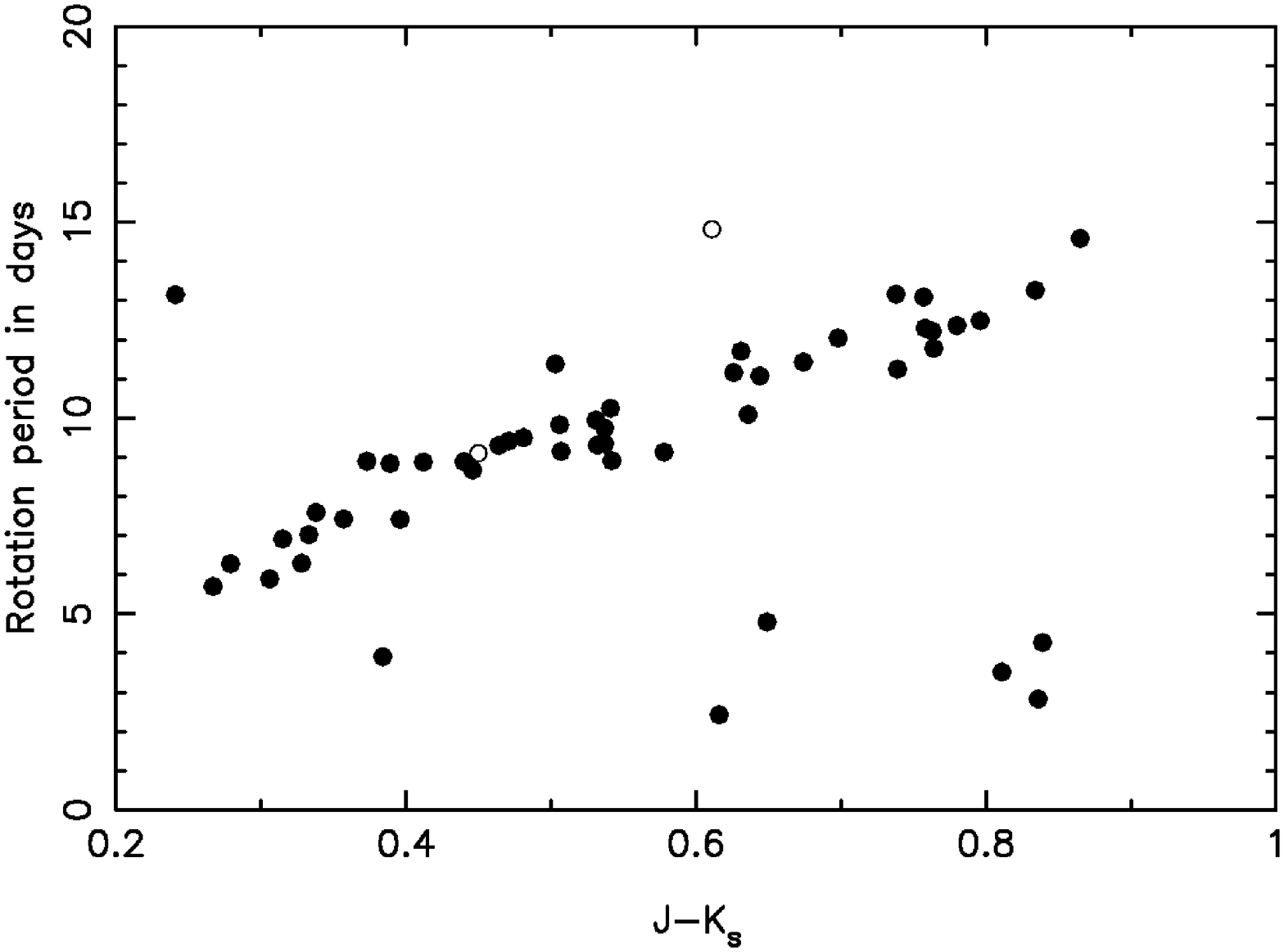} &
 \includegraphics[width=6.25cm,angle=270]{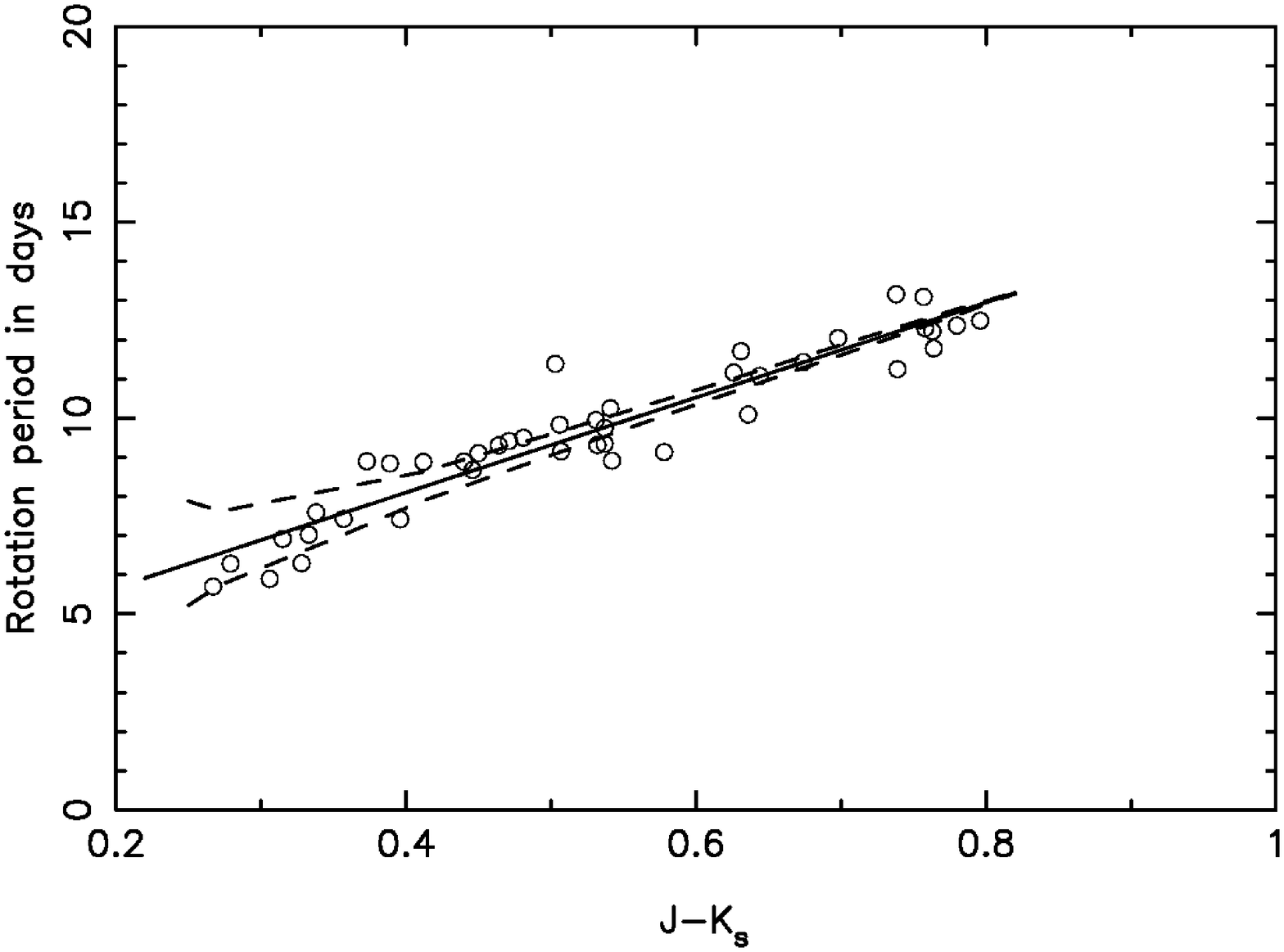} 
\end{tabular}
\caption{{\bf Left:}$J-K_s$ colour-period plot of all selected
  Praesepe members. {\bf Right:}$J-K_s$ colour-period plot of Praesepe members that are used
   to derive the colour-period relation (black line). The
   corresponding masses at 600Myr (in M$\odot$) are given below the colour scales. \label{col_perJK_P}}
\end{figure*}

  Fig.\ref{col_perJK_P} shows the cluster members retained to derive
  the following colour-period relation:
$$ 
P =   9.648 +   12.124 * (J-K_s -   0.528)
$$

\begin{figure}
 \includegraphics[width=6.25cm,angle=270]{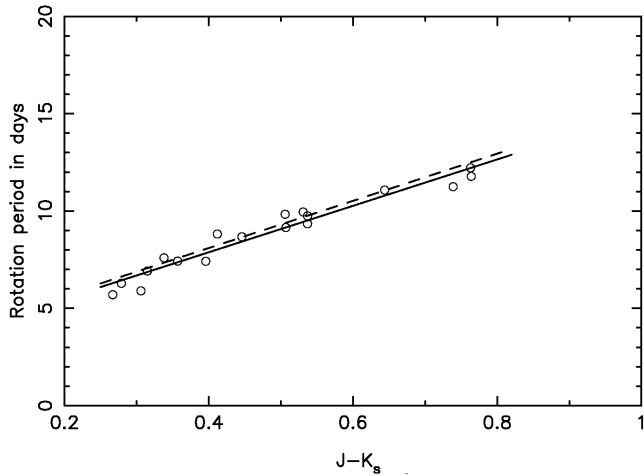}
 \caption{ Colour-period diagram and
   relation of a biased Praesepe sample obtained by manually setting a
   very tight constraint (sigma of 0.075 mag) on the colour-magnitude
   selection. Black line is the resulting colour-period relation. The
   regular colour-period relation obtained with the full sample is
   shown a dashed line for comparison.
   \label{P_tight}}
\end{figure}

 The standard deviation around the relation is 0.46day, again much
 higher than the measurement error on periods which is between 0.1
 and 0.2 day. It seems
 that for Praesepe as much as for the Hyades, the period-colour
 relation has not perfectly converged yet and is still slightly 
 dependent on the initial condition at the formation of the
 cluster. However, it should be again remarked that the scatter derived for the Coma
 cluster by CC09 is only 0.19day, while the cluster has about the same
 age as Praesepe and the Hyades. This could be a selection bias
   on CC09 Coma cluster member selection. Indeed Coma proper motion
   locus is within one sigma of the field proper motion distribution,
   and since the dispersion of the field proper motion is much larger
   than the dispersion of the cluster proper motions, objects close to
 both loci will systematically be classified as field objects rather
 than cluster members, unless the object's colour is close enough to the
 cluster main sequence to push the cluster membership
 probability over our p$>$0.5 threshold. This biased the Coma sample toward keeping only members
 which are very close to the cluster proper-motion and colour loci,
 resulting in less contamination and significant loss of
 completeness. Our study of Hyades and Praesepe uses the same cluster
 membership rationale but does not suffer from this bias because these
 cluster's proper motion locus are very far from the proper motion
 distribution of the field. We created a similar bias in
 Praesepe by  manually setting the sigma of the colour-magnitude
 distribution in the cluster membership calculation routines to a very
 small value (0.075mag) and obtained a similarly tight colour-period
 relation on the smaller number of candidates selected, see Fig
 \ref{P_tight}, with a scatter of 0.23. The number and dispersion of the
 selected objects is very similar to what is observed for Coma by CC09.
  Note that the same 
 procedure is not applicable to the Hyades because of the natural
 spread in magnitudes (see Fig. \ref{H_CMD}) induced by the large
 spatial extent of the cluster. 

 The astrophysical relation between
 selecting objects closer to the main-sequence at a given age and
 obtaining a tighter colour-period relation is not obvious and is
 likely down to a combination of several possible causes such as:
\begin{itemize}
\item Selecting only objects with magnitudes very close to the
  expected main sequence removes most of low to moderate contrast
  binaries. This eliminates the effect of unresolved binarity on
  periods and also 
  colours than can increase the period-colour scatter.
\item Objects whose colour or magnitude is affected by photometric
  measurement errors of any cause would be excluded. This directly
  removes most of the scatter in the period-colour relation which is
  due to colour measurement error.
\item Any object showing a noticeable colour or apparent magnitude
  variation from the main 
  sequence due to a higher or a lower than average spot coverage would
  be excluded from the member list. Since there is a strong
  correlation between rotation period and spot coverage this could
  also tend to select only objects with more average colour and rotation.
\item If there is any actual age spread within a cluster, a tighter
  colour-magnitude selection would eliminate older or younger objects,
  thus removing any scatter in the period-colour relation that could
  arise from age variations.
\item More generally, the colour-period relation actually stems from
a mass-period relation. Any factor such as metallicity or individual star
history, which would affect the colour-mass-magnitude relation  would
increase the scatter in the period-colour relation and would be
cut-out by tighter constraints on the colour-magnitude selection.
 \end{itemize}

 
\section{Comparison between the Hyades and Praesepe: age and braking.}
\subsection{Braking timescales}
\begin{figure*}
  \begin{tabular}{cc}
 \includegraphics[height=6.5cm,angle=0]{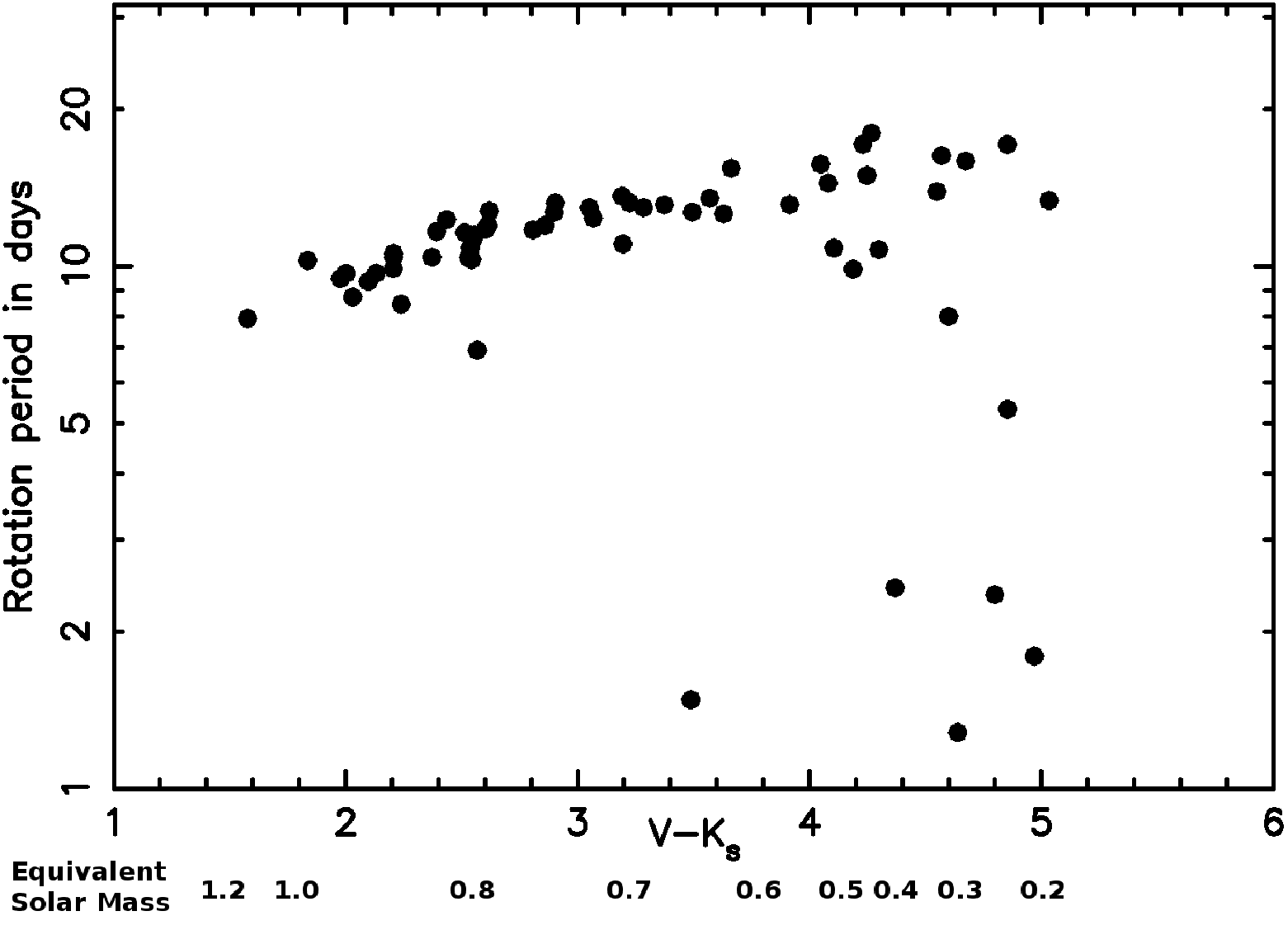} &
 \includegraphics[height=6.5cm,angle=0]{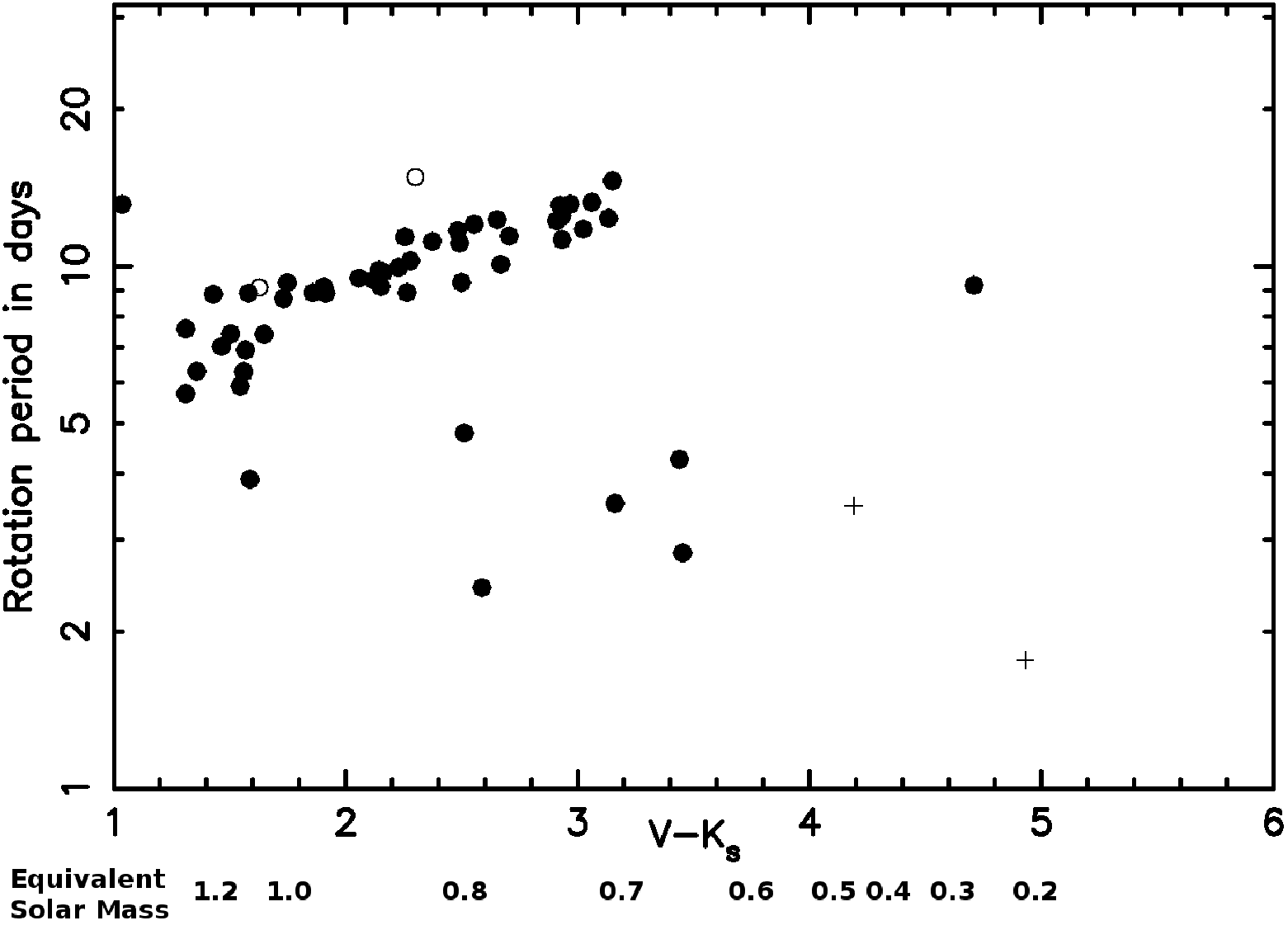}
 \end{tabular}
\caption{{\bf Left: }Log of rotation period-$V-K_s$ colour plot of
  objects identified as Hyades members. {\bf Right: }Same for Praesepe
  members. The crosses are 2 Praesepe objects from \citet{Scholz.2007}. Full
  circles are objects with a membership probability above 0.85, hollow
  circles are objects with a membership probability above 0.5.\label{col_perVK}}
\end{figure*}

 Deriving stellar rotation braking timescales over a wide
    range of stellar masses, and especially in the M-dwarf regime where
    stars become entirely convective, puts strong constraints on
 theoretical models of magnetic field/rotational braking.
As seen in Figs. \ref{col_perJK} and \ref{col_perJK_P}, our survey 
determined rotation periods of M-dwarf stars both in the Hyades and
Praesepe. However, these figures use the $J-K_s$ colour that tend to
saturate for the reddest stars ($J-K_s$ only varies from 0.82 to 0.86 when the stellar mass
varies from 0.6$M\odot$ to 0.25$M\odot$) that
are the most interesting to explore rotational braking
timescales. Fig. \ref{col_perVK}, shows the logarithmic rotation period as a function of
the $V-K_s$ colour which  varies from 3.7 to 4.8 as the stellar mass
varies from 0.6$M\odot$ to 0.25$M\odot$. Given this more than 40 times better colour
dynamic for low-mass objects,  using the $V-K_s$
colour allows a much more accurate determination of the stellar mass
range where the tight colour-period relation breaks down. However, a few of the stars
visible in the $J-K_s$ plots do not appear here, because they have no
$V-$band magnitudes.  Also, errors on $V$-band magnitudes are much
bigger than error on infrared magnitudes, especially since most of the
optical magnitudes are taken from photographic plates.\\ 

For the Hyades this breakdown is obvious because we have a good sampling at
low masses: the cluster is nearby and its M-dwarfs are bright
enough to derive reliable rotation periods. It seems to occur for
$V-K_s>$4.0 (i.e. masses $\sim < 0.5M\odot$), with the apparition of
numerous fast rotators as well as a significant increase of the slow
rotators scattering around the colour-period relation. However,
several slow rotators with periods in good agreement with the colour-period
relation are still observed down to the lowest masses sampled by our
survey, around  $0.2M\odot$. The Hyades data therefore demonstrate
that FGK and M stars above 0.5$M\odot$  have converged toward a
 simple colour-period relation by
Hyades age, about 625 Myr.  

The breakdown of the period-colour relation is less clear-cut for Praesepe because this
cluster is farther away and the survey detection limit is in the late K/early
M-dwarf range. It is however clear that objects bluer than
$V-K_s$=3.2 (or above 0.65M$\odot$) have converged toward a
the colour-period relation by Praesepe age. Three fast rotators
(1SWASP J084044.25+202818.7, 1SWASP J084245.96+211616.3 and  1SWASP
J084344.72+211234.5, the 2 latter being
independently identified during two seasons at the same period) are
however detected 
for  $V-K_s$=3.1-3.4, respectively 3.4, 3.1 and 3.4. This corresponds
to a mass of $\sim 0.65M\odot$ in
the late K-dwarf range,  where just one Hyades fast rotator (a
spectroscopic binary) is
identified. However, the $V$ magnitude of these faint red Praesepe
members is not very accurate and they are possibly later type objects,
as hinted by their $J-Ks$ colours,
0.839, 0.811 and 0.836 which would put all but one in the M dwarf range.
Spectroscopy from \citet{Allen.1995}
gives spectral types of K6, K7.5 and M0.8, which would put the last
one, 1SWASP J084344.72+211234.5 outside the mass range where other Praesepe
members have converged toward a clean colour-period
relation. This leaves 1SWASP J084044.25+202818.7 and 1SWASP
J084245.96+211616.3 as unexpected fast-rotating Praesepe members. \\ 
 If these two objects are representative
of the actual late K-dwarf cluster population,  
that would mean that Praesepe is significantly younger
than the Hyades, even more than the 50Myr we derive in the next section. Indeed, since all objects observed in this mass range in the
Hyades have spun down to 11-13days period, the age difference between
the two clusters must be significant enough to allow these Praesepe fast rotators
to spin down from their current 3.5-5day period to
11-13day when they will be of Hyades age. \\
 Assuming a naive constant braking
rate would have these objects rotating well beyond their break-up
velocities when they were 300Myrs old or younger. Since the braking
rate is supposed to decrease with time, it is even more incompatible
with wind-driven angular momentum loss in just 50Myr.
However an age difference significantly larger than that would be surprising
because Praesepe and the Hyades have been found to be of similar
age.\\
The alternative is that these two late K Praesepe fast rotators
would be short period 
binaries \citep[$<$12 days according to][]{Meibom.2005,Mazeh.2008,Raghavan.2010}, which are
not expected to follow the colour-period relation 
because of tidal interactions. A few percent of solar-type stars are
expected to be short period binaries \citep{Abt.1976,Duquennoy.1991},
making plausible that these Praesepe fast rotators are
such.
 They are not known as short period binaries, but they
have not surveyed for this. 1SWASP J084044.25+202818.7 appears
significantly over-luminous in a colour-magnitude diagram and at least
this one is likely an equal mass binary. 
It is therefore possible that both of them are close binaries and
would be simpler to account than a major age difference between the
two clusters, but this would need to be confirmed. \\
 \citet{Scholz.2007} have probed the mid/late M-dwarf rotation
periods in Praesepe and found only fast rotators. We added the 2
objects from their study which 
have $V$-band NOMAD magnitudes (the most massive of the 5 they
studied) in Fig. \ref{col_perVK}. Together with
our SWASP data this shows that Praesepe stars rotation rate have
clearly converged
for masses higher than 0.65$M\odot$ and have not converged in the
mid/late M dwarf range. The
colour-period relation 
breakdown in Praesepe must then occur in between, in the
0.65-0.4$M\odot$ mass range. The picture is
still unclear in this range because there is no significant
overlap between \citet{Scholz.2007} and our survey. A more accurate
determination of the colour-period relation in Praesepe would need
more data on early M dwarfs rotation periods in the cluster and/or to
ascertain whether or not the 3 red fast rotators we identified are
spectroscopic binaries. \\


\subsection{Gyrochronological ages of the Hyades and Praesepe}
\citet{Barnes.2003} coined the word ``gyrochronology" to describe the
technique that permits us to derive the age of a star when its rotation
period is known. This assumes, following the early study of
\citet{Skumanich.1972}, that the rotation period of stars of given
mass converges after a certain time to the same value independently of
the initial conditions and that their rotation period then evolves
following  a simple spindown law where $P\propto t^{b}$, with b=0.5
defined as the magnetic braking index. The data presented here
confirms that Hyades and Praesepe stars with 0.2$<J-K_s<$0.82 have already
converged toward a well defined colour-rotation period relation.

\subsubsection{Intra-cluster age distribution}
 Any
star in this colour range can have its period, and thus its age,
compared to the average period of the clusters, as defined by the
colour period relation of each cluster . We
can therefore use their rotation periods to accurately compute the relative age distribution
within each cluster. The age of an individual star relative to 
the fiducial rotational age $t_{\rm cluster}$ 
of the each population is given by: \\

\begin{equation}
t=t_{\rm Hyades}\left(\frac{P}{11.401 + 12.652(J-K_s-0.631)}\right)^2.
\label{eq:gyro_age1a}
\end{equation}
\begin{equation}
t=t_{\rm Praesepe}\left(\frac{P}{9.648 + 12.124(J-K_s-0.528)}\right)^2.
\label{eq:gyro_age1a}
\end{equation}

\begin{figure}
 \includegraphics[width=6.25cm,angle=270]{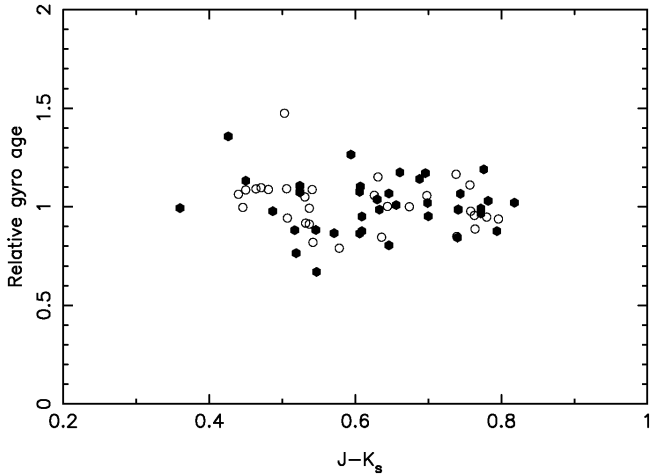} 
 \caption{Relative age of individual Hyades members (black circles) and
   Praesepe members (hollow circle) for slow rotators with 0.2$<J-K_s<$0.82,
   assuming an arbitrary age of 1 for both clusters.\label{agerel}}
\end{figure} 

 Setting the cluster ages to the arbitrary value of 1, we obtain an
average age for the Hyades sample of 1.006, with a 
dispersion of  0.14 and an average of  1.003 for the Praesepe
sample, with a dispersion of  0.10 (See Fig.\ref{agerel}). Note that
these are relative ages within each cluster which are meant to
illustrate the dispersion in age measurements within each cluster
  (which stems from 
  the dispersion around the period-colour relation and is not
  necessarily an actual age dispersion) and are not relevant to
compare the age of the cluster themselves.  Under the assumption that
all stars in 
each cluster have about the same age, the dispersions of 10 and 14\% we
derive here are representative of the accuracy of gyrochronology to
measure the age of individual stars.

\subsubsection{Relative age of Praesepe compared to the Hyades}
 To derive Praesepe's age
relatively to the Hyades, we anchored the age-period relation assuming
a mean Hyades age of 625 Myr (P98). We derived the age of Praesepe stars by
computing the 
rotation period they would have if they had the age of Hyades and comparing
this hypothetical period to the measured one we were able to compute
an age for each slow rotator with 0.2$<J-K_s<$0.82 in Praesepe as follows:
\begin{equation}
t=625\left(\frac{P}{11.401 + 12.652(J-K-0.631)}\right)^2.
\label{eq:gyro_age2}
\end{equation}

\begin{figure}
 \includegraphics[width=6.25cm,angle=270]{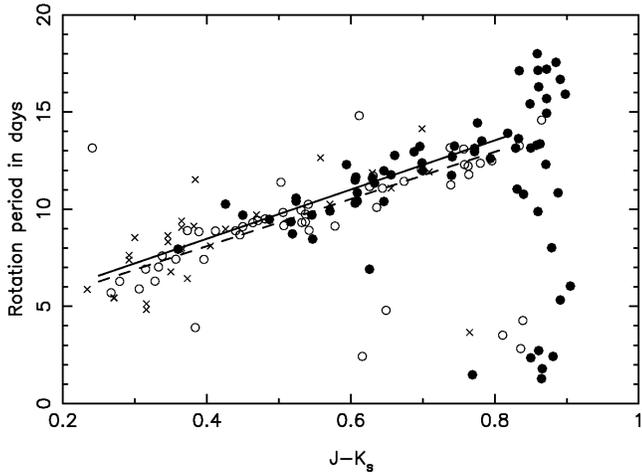} 
 \caption{$J-K_s$ colour-period plot of Hyades (black circles for
   SWASP objects and crosses for R87-95 objects) and
   Praesepe members (hollow circle). The period-colour relation are
   also shown, black line for the Hyades and dashed line for Praesepe.\label{col_perHP}}
\end{figure}

The age of Praesepe was derived from the averaged ages of these 43
stars, and stands at  573$\pm$13Myr. We carried out the same analysis
for the Hyades.  The average age of the stars in the Hyades
  sample is
628$\pm$14Myr.  These absolutes values need to be taken with caution
since the uncertainties on the actual Hyades age, that we use to
anchor the relation, are much bigger
\citep[$\pm$50Myr][]{Perryman.1998} than our errors. However, the
relative measurements 
are much more reliable and a 
Student $t$-test on these 0.2$<J-K_s<$0.82 samples showed that there is only 1.5\%
likelihood they derive from the same age distribution. This means it is highly
probable that Praesepe is actually younger than the Hyades, our
results pointing toward a 50Myr age difference. These quantitative
results are backed by the qualitative appearance of the clusters'
colour-period plots, Praesepe stars having shorter periods 
in average(See Fig.\ref{col_perHP}). As seen in the previous section,
this might also be supported by the presence of bluer fast rotators in
Praesepe than in the Hyades.

Using our new SWASP  data for the Hyades and CC09 data for the Coma
cluster, we also 
derived an improved estimation of Coma's age: 580$\pm$12Myr. This
agrees well with the 591$\pm$41Myr found by CC09 using older
Hyades data (from R87-95) to calibrate the cluster 's age and puts
Coma almost exactly at the same age as Praesepe. The ages derived with
the improved gyrochronological relation described in the last section of this
article are quite similar, with an age of
578$\pm$12Myr for Praesepe and 584$\pm$10Myr for Coma.

\subsubsection{Individual stars age distribution relative to Hyades age}
\begin{figure}
 \includegraphics[width=6.25cm,angle=270]{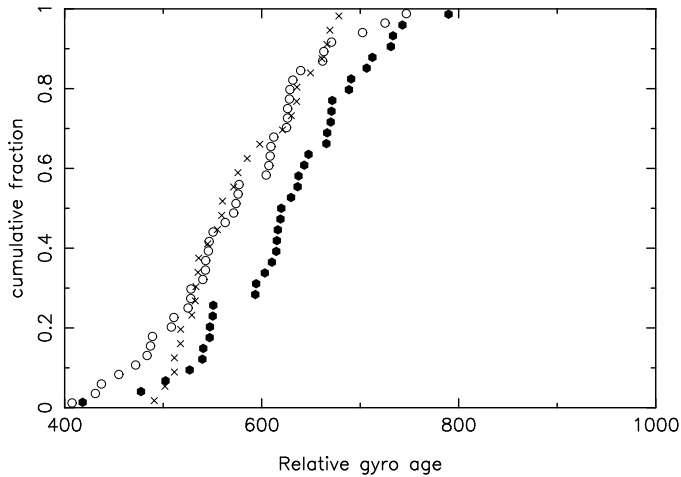} 
 \caption{Cumulative fraction plot of Hyades members ages (black circles),
   Praesepe members ages (hollow circles) and Coma members ages (crosses).\label{agecompare}}
\end{figure}

 Fig. \ref{agecompare}
shows the ages of individual stars in our sample for each of these 3
clusters. The large scatter in ages observed on this figure is 
likely not real but most of it is probably due to the scatter of the
colour-period-age relation at about 600Myr. This scatter is 85Myr for
the Hyades, 85Myr for Praesepe and 61Myr for Coma (respectively 76Myr,
77Myr and 55Myr with
the improved gyrochronological relation), showing that at
these ages the simple gyrochronology spin-down law from
\citet{Skumanich.1972} enables age measurements for individual stars with a
better than 15\% accuracy, improving as the square root of the number
of cluster members when measuring the age of a cluster. If this relation is
properly calibrated for field stars, gyrochronology should provide
even more accurate age measurements for individual field stars because
the scatter of the colour-period-age relation is expected to decrease
with age. This excellent precision in relative age measurement would be
extremely valuable even if it is somewhat degraded by systematic
errors arising from calibration of the technic. 

\section{Calibration of gyrochronolgy: the age of field stars}
\subsection{Combining all SWASP cluster data into an homogeneous set.}
  In order to calibrate gyrochronology using all the SWASP clusters
  data, we numerically aged the rotation periods of Praesepe and Coma
  stars from CC09 up to the Hyades age of 625Myr. We assumed a Skumanich braking
  law to derive this $P_{Hyades}$:
$$
P_{Hyades}=P_{Cluster}*(\frac{Age_{Hyades}}{Age_{Cluster}})^{1/2}
$$

\begin{figure}
 \includegraphics[width=6.25cm,angle=270]{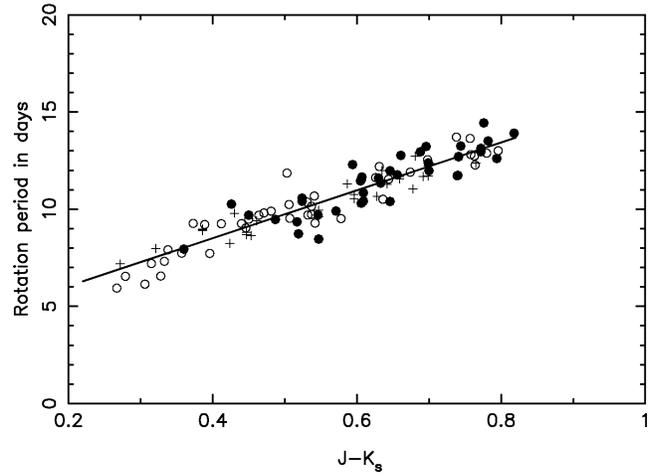} 
 \caption{Period-colour relation for Hyades members ages (black circles),
   artificially aged Praesepe members ages (hollow circles) and  artificially aged Coma members ages (crosses).\label{col_per_all}}
\end{figure}

We then fitted a simple linear relation through this combined data set
of all 109 cluster stars with SWASP periods, in the same way we
obtained the individual cluster period-colour relation. This gave the
following period-colour relation, that we used to
derive the age of field stars:

$$ 
P =   10.603 +   12.314 * (J-K_s -   0.570)
$$
The resulting dispersion around the relation is 0.45day. This compares
remarkably well with the weighted average of the dispersion around the
3 clusters 
colour-period relation, which stands at 0.44day: the artificial
``ageing" of the rotation periods introduces almost no noise in the
combined data. Fig. \ref{col_per_all} shows the periods and colour of
stars in this combined cluster data set and the relation fitted
through.

\subsection{The age of field stars: comparison with asterosismology}
 The period-colour relation from the combined SWASP cluster dataset
 gives us a statistically robust estimates of the rotation of FGK and
 early M stars by Hyades ages. We can derive the age of the Sun given
 its rotation period using the following relation:

\begin{equation}
Age_{Sun}=Age_{cluster}\left(\frac{Period(Sun)}{Period(1M\odot~star ~in ~cluster)}\right)^2.
\label{eq:gyro_age2}
\end{equation}
Using the isochrones from \citet{Pinsonneault.2004}, we have
$J-K_s$=0.399 for a 1M$\odot$ star in the Hyades, which translates
into a period of 8.5 day at Hyades ages. Assuming the Hyades ages is
625Myr, neglecting any metallicity effect, this relation with no
further calibration gives the Sun an age of 5870$\pm$630Myr, 2 $\sigma$
or about 30\% away from its measured age of 4570Myr (\ref{gyro_ages}). The error stated
here comes from the dispersion around the colour-period relation and assume the rotation period of the Sun at the mid-latitude sunspot belt
is 26.1days \citep{Donahue.1996} and has no errors.
 The same
calculation can be applied to any star with an accurate age
measurement and a known rotation period. Following CC09, we determine
the gyrochronological ages of the only 3 stars in the FGKM spectral
range with accurate age determination from asterosismology 
  assuming there is no systematic error in these ages derived from asterosismology.

Their $J-K$ colours at Hyades ages were determined using the masses of
the stars and the \citet{Pinsonneault.2004} isochrones. The mass of
$\alpha$CenA was taken as 1.1M$\odot$ and $\alpha$CenB as 0.91M$\odot$  following
\citet{Pourbaix.1999} and the mass of 70OphA as 0.89M$\odot$, from \citet{Eggenberger.2008}. The rotation
periods are from \citet{Donahue.1996} for the Sun, \citet{Barnes.2007}
for $\alpha$ Cen, and from \citet{Stimets.1980} and \citet{Noyes.1984} for
70 Oph A. The last row is based on the supposition that the true
period of 70 Oph A is twice the measured period, which may be a higher
frequency harmonic of the true period, as we observed for some cluster
stars. The 
asteroseismological ages for $\alpha$ Cen and 70 Oph A are from
\citet{Eggenberger.2004,Eggenberger.2008}. With the exception of the Sun, the
main source of error is the uncertainty on the stars' periods.
 The results using the standard magnetic braking index $b=0.5 $ are
 shown in the 4th column of table \ref{gyro_ages}. The ages found are
 systematically older than the asterosismological ages or the age of
 the sun, and are  about 30\% off, which is already as good or even
 better than most of the age measurement methods currently available. 

Since the \citet{Skumanich.1972} square root braking law derives from a
simple magnetic field/angular momentum coupling models, it is expected
that some of the parameters it neglects, such as for instance the
evolution of the momentum of inertia of a star as it ages, could
slightly modify this power law. We therefore calibrated the
gyrochronology relation by assuming a slightly different magnetic
braking index, $b=0.56$, already used by CC09. Using this modified law
 the period-colour relation we derived from the combined SWASP cluster
 dataset gives an almost exact age for the Sun, of 4.6$\pm$0.4Gyr. The
 resulting ages for the others stars are shown in the 5th column of table
 \ref{gyro_ages} and are also much closer to the asterosismological ages
 than those derived from the standard braking law. Note that assuming
 $b=0.56$ instead of $0.5$ does not noticeably affect the artificial
 ageing of Coma and Praesepe described above because these clusters
 have very similar ages.

 Another way to
 calibrate gyrochronology without modifying the magnetic
braking index would be to assume that the Hyades are younger than
625Myr. However we would need to assume that the Hyades are much
younger, 525Myr old, to obtain a solar age of 4.9$\pm$0.5 Gyr, and this would
still not fit the other stars asteroseismological ages within 1
sigma.

 We therefore conclude that calibrating gyrochronology by
assuming a modified magnetic braking index with $b=$0.56 provides a
much better fit of the ages of the sample of stars considered
here. Given the small age difference between Coma, Praesepe and the
Hyades this new calibration induce a less than 1\% difference in their
relative age. The corresponding ages are 578$\pm$12Myr for Praesepe
and 584$\pm$10Myr for Coma.\\

 \citet{Barnes.2007} already demonstrated that gyrochronology produced
much more self-consistent age measurements than chromospheric activity and
 isochrone fitting. When period
   measurement is accurate, (which is only the case of the Sun in
   table \ref{gyro_ages}) we have shown in section 6.2.1 and 6.2.2
   that our age measurements by gyrochronology for individual
   stars have an accuracy of 10 to 15\%. Our results therefore confirm
   earlier forecasts by 
 CC09, \citet{Barnes.2007} and \citet{Mamajek.2008} that correctly
 calibrated gyrochronological ages could give absolute age
 measurements with an accuracy of around 10\% for field stars in the
 FGK and early M spectral range. Note that this age-rotation
   period cannot be applied to derive a quantitative measurement of
   the age of stars younger than their magnetic braking convergence
   time, shown here to be $\sim$600Myrs for late K and early M-dwarfs.
  Hotter stars converge earlier, and F and G dwarfs have 
   converged toward a clean colour-period relation 
   at the age of M35 \citep[150Myr,][]{Meibom.2009_M35}. Achieving this
 accuracy would only 
 necessitate measuring the period of the star with an accuracy
 comparable to SWASP's. However these results
 need to be confirmed by comparing with a larger sample of stars
 with independent and accurate age measurements. To our knowledge
 there are no such objects beyond those discussed here.

\begin{table*}
\caption{Gyrochronological ages for the Sun and old main-sequence stars with measured rotation periods and asteroseismological age determinations.  }
\label{gyro_ages}
\begin{tabular}{lccccc}
\hline\\
Star & $J-K$ & Period & GyroAge & GyroAge & Seismo\\
     & at Hyades age & (days) & $(b=0.5)$ & $(b=0.56)$ & Age(Gyr)\\
\hline\\
Sun & 0.399 & $26.1\pm0.0$ & $5.9\pm0.6$ & $4.6\pm0.4$ & 4.57\\
$\alpha$CenA & 0.33 & $28.0\pm3.0$ & $8.5\pm2.1$ & $6.4\pm1.4$ & $6.5\pm0.3$\\
$\alpha$CenB & 0.49 & $36.9\pm1.8$ & $9.2\pm1.2$ & $6.9\pm0.8$ & $6.5\pm0.3$\\
70OphA & 0.51 & $19.9\pm0.5$ & $2.5\pm0.3$ & $2.2\pm0.2$ & $6.2\pm1.0$\\
70OphA & 0.51 & $39.8\pm1.0$ & $10.2\pm1.1$ & $7.5\pm0.7$ & $6.2\pm1.0$\\
\hline\\
\end{tabular}
\end{table*}

\section{Conclusion}
 We presented an analysis of SWASP data that found more than 120
 rotational variables that we identified as Hyades and Praesepe
 cluster members. This
 allowed us to put strong constraints on the rotational braking time
 by showing that the periods of all FGK and M single stars down to
 $\sim$0.5M$\odot$ in our sample have converged
 toward a relatively tight period-colour relation by Hyades age. We
 used gyrochronological relations and the period-colour relations
 derived for each cluster to accurately measure their relative ages,
 assuming the Hyades are 625Myr old. This yields ages of 578$\pm$12Myr
 for Praesepe and 584$\pm$10Myr for Coma and gives a statistically
 strong statement that Praesepe and the Hyades are not exactly co-eval
 and that the former is 47$\pm$17Myr younger than the latter. 
 We finally used the combined SWASP data using the Hyades stars, and
 artificially aged Praesepe and Coma stars to derive a global
 period-colour relation at Hyades age. We used it to calibrate the
 gyrochronology age relation with a modified magnetic braking index
 that fits well the age of the Sun and of the few stars with accurate
 asterosismological ages. This relation already gives reliable ages
 and hints that when properly calibrated,
 gyrochronology could be used to measure the absolute age for any FGK and early-M single field stars with an accuracy of about 10\%, provided their
 rotation period is measured.

\section*{Acknowledgments}

The WASP Consortium consists of representatives from the Universities of
Keele, Leicester, The Open University,Queens University Belfast and St
Andrews, along with the Isaac Newton Group (La Palma) and the
Instituto de Astrofisica de Canarias (Tenerife). The SuperWASP and WASP-S Cameras were constructed and operated with funds made available from
Consortium Universities and PPARC/STFC. This publication makes use of
data products from the Two Micron All Sky Survey,  which is a joint
project of the University of Massachusetts and the Infrared Processing
and  Analysis Censer/California Institute of Technology, funded by the
National Aeronautics and  Space Administration and the National
Science Foundation. This research has made use of the VizieR catalogue
access tool, CDS, Strasbourg, France.

\bibliographystyle{aa}
\bibliography{bib}

\appendix
\begin{table*}
\caption{Cross-identifications and $B, V$ photometry from the
  litterature of  rotational variables with Hyades membership probabilities
  greater than 
  0.5. The 1SWASP identifier encodes the J2000.0 coordinates of the
  object. References are as follows, 1:\citet{Upgren.1979}, 2:\citet{Weis.1983},, 3:\citet{Upgren.1985}, 4:\citet{Weis.1988}, 5:\citet{Stauffer.1982}
}
\label{AnnexHyades}
\begin{tabular}{ccccccc}
\hline\\
1SWASP	& USNO B-1.0	& Cluster member Id &  
Spectral Type& $V$ & $B-V$ & Reference\\
\hline\\
1SWASPJ033734.97+212035.4	& 1113-0041831	& Melotte 25 5 	& G5	& 9.36	& 0.92	& 2		\\
1SWASPJ034347.07+205136.4	& 1108-0041451	& -	& -	& 14.54	& 9	& 4	                \\
1SWASPJ035234.31+111538.6	& 1012-0033507	& -	& M	& 13.73	& 1.54	& 2	                \\
1SWASPJ035453.20+161856.3	& 1063-0039925	& -	& M	& 14.25	& 1.58	& 2                     \\
1SWASPJ035501.43+122908.1	& 1024-0045401	& Melotte 25 S 155 	& M0	& 10.12	& 1.07	& 1     \\
1SWASPJ040339.03+192718.1	& 1094-0044958	& Melotte 25 222	& K5	& 10.17	& 1.08	& 2     \\
1SWASPJ040525.67+192631.7	& 1094-0045433	& Melotte 25 226	& K7	& 11.41	& 1.35	& 1     \\
1SWASPJ040634.62+133256.8	& 1035-0040066	& Melotte 25 REID 62	& M	& 13.52	& 1.47	& 9     \\
1SWASPJ040701.22+152006.0	& 1053-0044486	& Melotte 25 231	& K5	& 10.49	& 1.18	& 1     \\
1SWASPJ040743.19+163107.6	& 1065-0042472	& Melotte 25 229	& K0	& 9.94	& 1.02	& 2     \\
1SWASPJ040811.02+165223.3	& 1068-0041870	& Melotte 25 REID 72 	& K7	& 11.52	& 1.44	& 1     \\
1SWASPJ040826.66+121130.6	& 1021-0043331	& Melotte 25 233	& M0	& 11.28	& 1.33	& 2     \\
1SWASPJ040836.21+234607.0	& 1137-0048046	& Melotte 25 PELS 18 	& G5	& 9.44	& 0.9	& 1     \\
1SWASPJ041127.64+155931.8	& 1059-0051230	& Melotte 25 HAN 43	& M	& 15.15	& 1.65	& 2     \\
1SWASPJ041510.33+142354.5	& 1043-0041060	& Melotte 25 245	& K2	& 11.54	& 1.38	& 2     \\
1SWASPJ041633.47+215426.8	& 1119-0053591	& Melotte 25 21	& G5	& 9.14	& 0.81	& 2             \\
1SWASPJ041725.14+190147.4	& 1090-0048172	& Melotte 25 251	& K5	& 10.83	& 1.22	& 2     \\
1SWASPJ041728.13+145403.9	& 1049-0042422	& Melotte 25 VA 106	& M3	& 14.46	& 1.55	& 5     \\
1SWASPJ042322.85+193931.1	& 1096-0050945	& Melotte 25 43	& K2	& 9.4	& 0.91	& 2             \\
1SWASPJ042325.28+154547.2	& 1057-0062216	& Melotte 25 173	& K0	& 10.49	& 1.24	& 1     \\
1SWASPJ042350.70+091219.5	& 992-0039564	& -	& M3	& 12.89	& 1.51	& 1                     \\
1SWASPJ042359.13+164317.7	& 1067-0045253	& Melotte 25 260	& M0.5	& 12.55	& 1.49	& 3     \\
1SWASPJ042416.93+180010.4	& 1080-0063868	& Melotte 25 174	& K5	& 9.99	& 1.06	& 1     \\
1SWASPJ042500.18+165905.8	& 1069-0045244	& Melotte 25 175	& K4	& 10.3	& 1.03	& 1     \\
1SWASPJ042514.54+185824.9	& 1089-0051192	& Melotte 25 PELS 40	& 	& 12.82	& 1.48	& 1     \\
1SWASPJ042547.55+180102.2	& 1080-0064462	& Melotte 25 176	& K2	& 9.01	& 0.94	& 1     \\
1SWASPJ042642.81+124111.7	& 1026-0052108	& Melotte 25 269	& M0	& 10.48	& 1.36	& 2     \\
1SWASPJ042648.25+105215.9	& 1008-0039886	& Melotte 25 271	& K5	& 9.45	& 1.04	& 2     \\
1SWASPJ042725.34+141538.3	& 1042-0043123	& Melotte 25 177	& K2	& 10.3	& 1.08	& 4     \\
1SWASPJ042747.03+142503.8	& 1044-0042884	& Melotte 25 179	& K0	& 9.49	& 0.92	& 3     \\
1SWASPJ042828.78+174145.1	& 1076-0062237	& Melotte 25 VA 486 	& M1	& 12.05	& 1.49	& 5     \\
1SWASPJ043033.88+144453.1	& 1047-0044652	& Melotte 25 LH 75	& M	& 14.68	& 1.56	& 5     \\
1SWASPJ043034.87+154402.3	& 1057-0063658	& Melotte 25 182	& K0	& 8.93	& 0.84	& 1     \\
1SWASPJ043152.40+152958.3	& 1054-0052709	& Melotte 25 191	& K0	& 11.04	& 1.31	& 1     \\
1SWASPJ043225.66+130647.6	& 1031-0059808	& Melotte 25 288	& K0	& 10.91	& 1.19	& 2     \\
1SWASPJ043323.75+235927.0	& 1139-0053890	& Melotte 25 PELS 74 	& K6	& 12.62	& 1.52	& 1     \\
1SWASPJ043327.00+130243.6	& 1030-0056878	& -	& -	& 13.16	& 1.57	& 1                     \\
1SWASPJ043337.18+210903.0	& 1111-0054738	& Melotte 25 290	& K0	& 10.7	& 1.23	& 2     \\
1SWASPJ043411.14+113328.4	& 1015-0040520	& Melotte 25 294	& K4	& 11.73	& 1.39	& 2     \\
1SWASPJ043548.51+131717.0	& 1032-0063919	& Melotte 25 VA 776	& M2.5	& 14.91	& 1.63	& 1     \\
1SWASPJ043605.26+154102.4	& 1056-0061021	& Melotte 25 99	& K0	& 9.37	& 0.87	& 2             \\
1SWASPJ043950.97+124342.5	& 1027-0058818	& Melotte 25 311	& K5	& 10.06	& 1.07	& 2     \\
1SWASPJ044127.81+140434.1	& 1040-0045181	& -	& M2.5	& -	& -	& -                     \\
1SWASPJ044128.75+120033.7	& 1020-0044793	& -	& -	& 12.88	& 1.5	& 4                     \\
1SWASPJ044129.67+131316.3	& 1032-0065593	& Melotte 25 316	& 	& 11.23	& 1.44	& 3     \\
1SWASPJ044142.98+082620.0	& 984-0050517	& -	& -	& -	& -	& -                     \\
1SWASPJ044315.69+170408.7	& 1070-0053038	& Melotte 25 319	& K0	& 9.86	& 1	& 2     \\
1SWASPJ044618.79+033810.7	& 936-0059947	& Melotte 25 326	& K4.5	& 10.93	& 1.27	& 2     \\
1SWASPJ044630.38+152819.3	& 1054-0056655	& Melotte 25 142	& G5	& 8.3	& 0.66	& 2     \\
1SWASPJ044718.51+062711.6	& 964-0047473	& -	& K6	& 11.33	& 1.42	& 2                     \\
1SWASPJ044800.88+170321.6	& 1070-0054253	& Melotte 25 331	& 	& 11.12	& 1.41	& 4     \\
1SWASPJ044830.61+162319.0	& 1063-0051391	& -	& -	& 12.42	& 1.47	& 4                     \\
1SWASPJ044842.13+210603.6	& 1111-0060275	& Melotte 25 115	& G5	& 9.06	& 0.85	& 2\\
1SWASPJ044912.98+244810.2	& 1148-0059776	& Melotte 25 117	& K3	& 9.53	& 1.04	& 4\\
1SWASPJ044952.11+060633.6	& 961-0047586	& -	& -	& 14.73	& 1.63	& 4\\
1SWASPJ045000.70+162443.5	& 1064-0050763	& -	& K0	& 10.61	& 1.16	& 1\\
1SWASPJ045102.41+145816.5	& 1049-0049205	& -	& M0	& -	& -	& -\\
1SWASPJ045223.53+185948.9	& 1089-0058167	& Melotte 25 PELS 98 	& K0	& 10.29	& 1.07	& 1\\
1SWASPJ045223.86+104309.9	& 1007-0045668	& -	& -	& -	& -	& -\\
1SWASPJ050540.37+062754.6	& 964-0053308	& Melotte 25 151	& K2	& 9.92	& 0.95	& 2\\
1SWASPJ051109.69+154857.5	& 1058-0068980	& -	& K4.5	& -	& -	& -\\
1SWASPJ051119.30+075432.0	& 979-0066501	& -	& K5	& -	& -	& -\\
\hline\\
\end{tabular}
\end{table*}

\begin{table*}
\caption{Cross-identifications and $B, V$ photometry from the
  litterature of  rotational variables with Praesepe membership probabilities
  greater than 
  0.5. The 1SWASP identifier encodes the J2000.0 coordinates of the
  object. References are as follows, 1:\citet{Klein-Wassink.1927}, 2:\citet{Jones.1991}, 3:\citet{Johnson.1952}, 4:\citet{Upgren.1979}, 5:\citet{Weis.1981}, 6:\citet{Mermilliod.1990}
}
\label{AnnexPraesepe}
\begin{tabular}{ccccccc}
\hline\\
1SWASP	&	USNO B-1.0	&	Cluster member Id &  
Spectral Type& $V$ & $B-V$ & Reference\\
\hline\\
J083141.93+260641.0	& 1161-0156334	& -	& -	& -	& -	& -\\
J083554.98+224611.3	& 1127-0193591	& -	& -	& -	& -	& -\\
J083622.69+191129.3	& 1091-0164534	& JS127	& K2	& 12.41	& 1.11	& 2\\
J083627.81+175453.4	& 1079-0203318	& JS134	& F7.5	& 10.96	& 0.37	& 2\\
J083627.86+210716.2	& 1111-0173917	& JS131	& K5.9	& 14.04	& -	& 2\\
J083628.30+201342.8	& 1102-0162570	& KW535	& K1.3	& 12.65	& 0.95	& 5\\
J083648.96+191526.4	& 1092-0164565	& KW539	& G4	& 10.76	& -	& 1\\
J083657.82+213355.9	& 1115-0175522	& JS156	& K7.7	& 14.46	& -	& 2\\
J083711.48+194813.2	& 1098-0164221	& KW23	& G	& 11.33	& 0.71	& 3\\
J083718.29+194156.3	& 1096-0163761	& KW27	& G5	& 11.45	& 0.74	& 3\\
J083722.23+201037.0	& 1101-0161763	& KW30	& G	& 11.4	& 0.72	& 3\\
J083727.54+193703.1	& 1096-0163804	& KW32	& G8	& 11.65	& 0.77	& 3\\
J083733.07+183915.5	& 1086-0168273	& JS189	& G0	& 10.6	& 0.74	& 2\\
J083735.77+205927.5	& 1109-0174003	& JS190	& K3.5	& 12.83	& 1.11	& 2\\
J083746.41+193557.5	& 1095-0163684	& KW48	& G8.7	& 12.32	& 0.9	& 4\\
J083747.39+190624.8	& 1091-0164888	& KW52	& G8.6	& 12.28	& 0.91	& 4\\
J084036.23+213342.1	& 1115-0176177	& JS369	& G6.2	& 11.64	& -	& 2\\
J084245.96+211616.3	& 1112-0175801	& -	& K7.5	& -	& -	& -\\
J084344.72+211234.5	& 1112-0175989	& JS545	& M0.8	& 15.19	& -	& 2\\
J084624.31+170235.1	& 1070-0184311	& -	& -	& -	& -	& -\\
J084714.12+162347.3	& 1063-0162881	& -	& F5.7	& -	& -	& -\\
J084847.83+274136.5	& 1176-0212100	& -	& -	& -	& -	& -\\
J083857.23+201053.6	& 1101-0162189	& KW560	& K6	& 13.93	& 1.33	& 4\\
J083902.28+191934.4	& 1093-0163918	& KW141	& K	& 12.39	& 0.93	& 4\\
J083921.54+204529.4	& 1107-0171494	& JS286	& G7.6	& 11.86	& 0	& 2\\
J083930.44+200408.6	& 1100-0163760	& KW182	& F8	& 10.31	& 0.68	& 3\\
J083935.53+185236.8	& 1088-0166993	& KW196	& G	& 10.73	& 0.65	& 3\\
J083945.78+192201.1	& 1093-0164074	& KW208	& G0	& 10.66	& 0.58	& 3\\
J084005.72+190130.7	& 1090-0167104	& KW256	& K3.2	& 12.62	& 1	& 4\\
J084022.41+203827.1	& 1106-0171109	& JS353	& K7	& 14.61	& -	& 2\\
J084025.56+192832.8	& 1094-0163661	& KW293	& F4.8	& 9.85	& 0.47	& 3\\
J084033.46+193801.0	& 1096-0164636	& KW313	& K0.8	& 12.18	& 0.87	& 4\\
J084044.25+202818.7	& 1104-0163841	& JS379	& K6	& 13.93	& -	& 2\\
J084047.61+185411.9	& 1089-0167723	& KW336	& G6.8	& 11.45	& 0.74	& 3\\
J084048.34+195518.9	& 1099-0165066	& KW335	& G3	& 11.03	& 0.65	& 3\\
J084055.31+183459.0	& 1085-0167929	& JS398	& K7.4	& 14.3	& -	& 2\\
J084059.70+182204.5	& 1083-0186951	& JS402	& K2.2	& 11.91	& 1.1	& 2\\
J084122.57+185602.0	& 1089-0167811	& KW390	& K3	& 12.96	& 1.04	& 4\\
J084130.69+185218.6	& 1088-0167362	& KW401	& K3.1	& 12.97	& 1	& 5\\
J084143.67+195743.8	& 1099-0165295	& KW417	& K1.1	& 12.35	& 0.88	& 4\\
J084158.83+200627.2	& 1101-0162974	& JS466	& K4	& 13.23	& 1.11	& 5\\
J084220.09+190905.7	& 1091-0165816	& JS482	& K5	& 13.51	& 1.24	& 2\\
J084237.01+200832.0	& 1101-0163099	& JS493	& K6	& 13.99	& 1.27	& 5\\
J084248.47+203424.5	& 1105-0167705	& JS503	& K0.8	& 12.57	& 0.78	& 2\\
J084308.22+194247.7	& 1097-0164856	& JS520	& K4	& 13.33	& 1.15	& 5\\
J084317.83+203037.3	& 1105-0167843	& JS529	& K1.0	& 11.82	& 0.83	& 6\\
J084332.39+194437.8	& 1097-0164949	& KW514	& K0	& 12.34	& 0.93	& 4\\
J084519.17+190010.9	& 1090-0168262	& JS591	& K5.4	& 13.82	& 0.78	& 2\\
J084647.32+193840.9	& 1096-0165969	& JS638	& G8	& 10.75	& 0.6	& 6\\
J084817.25+165447.7	& 1069-0177337	& -	& -	& -	& -	& -\\
J084948.35+160750.7	& 1061-0165009	& -	& -	& -	& -	& -\\
J090222.37+182223.9	& 1083-0190841	& -	& -	& -	& -	& -\\
\hline\\
\end{tabular}
\end{table*}




\label{lastpage}

\end{document}